\documentclass[12pt,preprint]{aastex}

\shorttitle{Bolocam Lockman Hole Survey}
\shortauthors{Laurent, Aguirre, Glenn et al.}

\newcommand{\D}{\,d}

\newcommand{\vect}[1]{\ensuremath{\mathbf{#1}}}

\begin{document}


\title{The Bolocam Lockman Hole Millimeter-Wave Galaxy Survey: Galaxy
Candidates and Number Counts}


\author{
G.~T. Laurent\altaffilmark{1,5},
J.~E. Aguirre\altaffilmark{1},
J. Glenn\altaffilmark{1},
P.~A.~R. Ade\altaffilmark{2}, 
J.~J. Bock\altaffilmark{3,4},
S.~F. Edgington\altaffilmark{4},
{A. Goldin}\altaffilmark{3},
S.~R. Golwala\altaffilmark{4},
D. Haig\altaffilmark{2},
A.~E. Lange\altaffilmark{4},
P.~R. Maloney\altaffilmark{1},
P.~D. Mauskopf\altaffilmark{2},
H. Nguyen\altaffilmark{3},
P. Rossinot\altaffilmark{4},
J. Sayers\altaffilmark{4}, \&
P. Stover\altaffilmark{1}}

\altaffiltext{1}{Center for Astrophysics and Space Astronomy \& Department of Astrophysical and Planetary Sciences, University of Colorado,
    593 UCB, Boulder, CO 80309-0593}
\altaffiltext{2}{Physics and Astronomy, Cardiff University, 5 The Parade,
P.O. Box 913, Cardiff CF24 3YB, Wales, UK}
\altaffiltext{3}{Jet Propulsion Laboratory, California Institute of
Technology, 4800 Oak Grove Drive, Pasadena, CA 91109}
\altaffiltext{4}{California Institute of Technology, 1200 E. California
Boulevard, MC 59-33, Pasadena, CA 91125}
\altaffiltext{5}{glaurent@colorado.edu}







\begin{abstract}

We present results of a new deep 1.1 mm survey using Bolocam, a millimeter-wavelength bolometer array
camera designed for mapping large fields at fast scan rates, without chopping.  A map, galaxy candidate list, and derived number counts are presented.  This survey encompasses 324 
arcmin$^2$ to 
an rms noise level (filtered for point sources) of $\sigma_{1.1\mathrm{mm}} \simeq$ 1.4 mJy beam$^{-1}$ 
and includes the entire regions surveyed by the published 8 mJy 850 $\mu$m JCMT SCUBA and 1.2 mm IRAM MAMBO surveys.  We reduced the data using a custom software pipeline to remove 
correlated sky and instrument 
noise via a principal component analysis.  Extensive simulations and jackknife tests were performed to confirm the robustness of our source candidates and estimate the effects of false 
detections, bias, and completeness.  In total, 17 source candidates
were detected at a significance $\ge 3.0\ \sigma$, with six expected false detections.  Nine candidates are new detections, while eight candidates have coincident SCUBA 850 $\mu$m and/or 
MAMBO 1.2 mm 
detections.  From our observed number counts, we estimate the underlying differential number 
count distribution of submillimeter galaxies and find it to be in general agreement with previous surveys.  Modeling the spectral energy distributions of these submillimeter galaxies after observations of 
dusty nearby galaxies suggests extreme luminosities of $L = (1.0-1.6) \times 10^{13}$ $\mathrm{L}_\odot$ and, if powered by star formation, 
star formation rates of $500-800$ $\mathrm{M_\odot}\mathrm{yr}^{-1}$.

\end{abstract}





\section{Introduction}

Submillimeter galaxies are extremely luminous ($L > 10^{12} L_\odot$),
high-redshift ($z > 1$), dust-obscured galaxies detected by their thermal dust emission (for a review see Blain et
al.\ 2002).  The dust is heated by
the ultraviolet and optical flux from young stars associated with
prodigious inferred star formation rates (SFRs) of $\sim100 - 1000$
M$_\odot$ yr$^{-1}$ \citep{blain02}.  Although $\sim \frac{1}{3}$ of sources appear to
contain an active galactic nucleus \citep[AGN;][]{alexander03,ivison04}, in nearly all cases the AGNs
are not bolometrically important \citep[$<20$\%;][]{alexander04}. Given
these SFRs, a burst of duration $10^8$ yr would be sufficient to form
all the stars in an elliptical galaxy, making it plausible that submillimeter galaxies are
the progenitors of elliptical galaxies and spiral bulges
\citep{smail02, swinbank04}.  Deep (sub)millimeter surveys
with SCUBA \citep{holland99} and MAMBO \citep{bertoldi00} have now resolved
10\%$-$40\% of the cosmic far-infrared background
\citep{puget96,hauser98,fixsen98} into submillimeter galaxies in blank-field surveys \citep[e.g.,][]{greve04,borys03,
scott02} and 40\%$-$100\% using lensing galaxy
clusters \citep{blain99,cowie02}.  Photometric redshifts constrain most of
the submillimeter galaxies found so far to lie at $z > 1$ \citep[e.g.,][]{ 
carilli99,yun02,aretxaga03}.

Recent observations of submillimeter galaxies with the Keck Low-Resolution Imaging Spectrograph (LRIS; Chapman et al.\ 2003a, 2003b, 2005) have shown that redshifts can be
obtained for $\sim$70\% of bright submillimeter galaxies ($S_{\mathrm{850 \mu m}} > $~5 mJy)
with bright radio counterparts \citep[$S_{\mathrm{1.4 GHz}} > $~30 $\mu$Jy;][]{blain04} \citep[the fraction of submillimeter galaxies with radio                       
counterparts appears to be $\sim$65\%;][]{ivison02}.  This sample of
submillimeter galaxies lies in a distribution peaking at $z = 2.4$ with $\Delta z= 0.65$
\citep{ivison02,chapman03b,chapman05}.  Six of these galaxies have had their
optical spectroscopic redshifts confirmed by millimeter CO line
measurements \citep{frayer98,frayer99,neri03,sheth04}.  This redshift distribution may be biased with respect to the 
overall submillimeter galaxy population owing 
to a number of selection effects:  the requirement of precise radio positions prior to spectroscopy, which introduces a 
bias against cooler galaxies, especially at higher redshifts ($z > 2.5$); limited completeness ($\sim 30\%$) of the
spectroscopic observations, which biases the sample against galaxies with weak emission lines; and the redshift gap at $z 
= 1.2-1.8$ due to the ``spectroscopic desert,'' in which no strong rest-frame ultraviolet
lines are redshifted into the optical.



Bolocam is a new millimeter-wave bolometer camera for the Caltech
Submillimeter Observatory (CSO).\footnote{See http://www.cso.caltech.edu/bolocam.}  Bolocam's large field of view (8\arcmin),
31\arcsec\ beams (FWHM at $\lambda = 1.1$ mm), and AC biasing scheme make
it particularly well-suited to finding rare, bright submillimeter galaxies
and for probing large-scale structure.  We have used Bolocam to conduct a
survey toward the Lockman Hole for submillimeter galaxies.  The Lockman Hole is a region in UMa in which absorbing material, 
such as dust and galactic hydrogen, is highly rarefied \citep[H$_\mathrm{I}$ 
column density of $N_\mathrm{H} \approx 4.5 \mathrm{\ x\ } 10^{19}$ cm$^{-2}$;][]{jahoda90}, 
providing a transparent window for sensitive extragalactic surveys over a wide spectral range, from the infrared and 
millimeter wavebands to UV and X-ray observations.  Submillimeter \citep{scott02, fox02, ivison02, eales03}
and millimeter-wave \citep{greve04} surveys for submillimeter galaxies
have already been done toward the Lockman Hole.  It is one of the
one-quarter square degree fields of the SCUBA SHADES,\footnote{See http://www.roe.ac.uk/ifa/shades/links.html.} the focus of a deep extragalactic survey with {\it XMM-Newton} 
\citep{hasinger01}, and a 
target field for {\it Spitzer} guaranteed time observations.  
The coverage of the Lockman Hole region by several surveys therefore makes it an excellent field
for intercomparison of galaxy candidate lists and measuring spectral energy distributions (SEDs), which will ultimately enable dust temperatures and redshifts to be constrained.

This paper is arranged as follows.  In \S\ \ref{section:observations} Bolocam and the
observations are described.  In \S\ \ref{section:datareduction} the data reduction pipeline,
including pointing and flux calibration, cleaning and sky subtraction, and
mapping, is described. In \S\ \ref{section:sourcelist} the source candidate list and tests of
the robustness of the candidates are presented.  We devote \S\ \ref{section:simulations} to the extraction of the number counts versus flux density relation using simulations designed 
to 
characterize the systematic effects in the data
reduction and the false detections, completeness, and bias in the survey.
In \S\ \ref{section:discussion} we discuss the implications of the survey, in \S\ \ref{section:futurework} we describe future work for this program, and
in \S\ \ref{section:conclusions} we give conclusions.

\section{Observations}
\label{section:observations}


\subsection{Instrument Description}

The heart of Bolocam is an array of 144 silicon nitride micromesh (``spider-web'') bolometers cooled to 260 mK by a three-stage ($^4$He/$^3$He/$^3$He)
sorption refrigerator.  An array of close-packed ($1.5f\lambda$),
straight-walled conical feed horns terminating in cylindrical waveguides and
integrating cavities formed by a planar backshort couples the bolometers to
cryogenic and room-temperature optics.  The illumination on the
10.4 m diameter CSO primary mirror is
controlled by the combination of the feed horns and a cold (6 K) Lyot stop,
resulting in 31\arcsec\ beams (FWHM).  A stack of resonant metal-mesh
filters form the passband in conjunction with the waveguides.  A
$\lambda=2.1$ mm configuration is also available, which is used for
observations of the Sunyaev-Zel'dovich effect and secondary anisotropies in
the cosmic microwave background radiation.  Technical details of Bolocam
are given in \cite{glenn98, glenn03} and \cite{haig04}; numerical simulations of the integrating cavities are described
in \cite{glenn02}.

A key element of Bolocam is the bolometer bias and readout electronics: an
AC biasing scheme (130 Hz) with readout by lock-in amplifiers enables the detectors
to be biased well above the $1/f$ knee of the electronics.  The electronic readout stability, in conjunction with Bolocam's rigorous sky noise subtraction algorithm, eliminates the need to nutate
the CSO subreflector.  Another advantage of this AC biasing
scheme is that it is easy to monitor the bolometer operating voltage; these voltages are determined by the total
atmospheric emission in the telescope beam and the responsivity of the bolometers.  Thus, a voltage that is a monotonic function of the in-band atmospheric
optical depth and bolometer responsivity is continually measured.  Sky subtraction is
implemented by either a subtraction of the average bolometer voltages or a
principal component analysis (PCA) technique, which is described below.

\subsection{Scan Strategy}

Two sets of observations of the Lockman Hole East ($\mathrm{R.A.} = 10^\mathrm{h}\,52^\mathrm{m}\,08^\mathrm{s}.82$,
$\mathrm{decl.} = +57^\circ\,21'\,33\arcsec.80$, J2000.0) were made with Bolocam: 2003 January, when the data were
taken with a fast raster scan without chopping (hereafter referred to as
raster scan observations), and 2003 May, when the data were acquired with a
slow raster scan with chopping as a test of this observing mode (referred
to as chopped observations).  Approximately 82 hr of integration time
over 17 nights were obtained on the field during 2003 January, resulting in
259 separate observations, and 41 hr over 19 nights were obtained in 2003 May, resulting in 64 separate observations.  The weather was generally
good during the 2003 January run and mediocre to poor during the 2003 May run, where
we characterize the weather quality by the sky noise variability (rapid
variability of the optical depth).\footnote{Another weather measurement influencing the Bolocam mapping speed is the CSO 225 GHz heterodyne, narrowband, ``tipper tau'' monitor, which 
measures the zenith 
atmospheric attenuation.  The 2003 January 
and May Lockman Hole observations yielded $\tau_{225\mathrm{GHz}}$ ranges and 75th percentiles of $\tau_{225\mathrm{GHz}}=0.028-0.129$, $\tau_{75\%}=0.083$ and 
$\tau_{225\mathrm{GHz}}=0.014-0.307$, 
$\tau_{75\%}=0.200$, respectively.}  One 
hundred and nineteen bolometer
channels were operational during 2003 May; however, only 89 bolometer
channels were used in the analysis of the 2003 January data.  (The remaining bolometers were not included in making the final Lockman Hole map because of excess noise and/or 
electronics 
failures.)
Nominally one feed horn per hextant is blocked to enable these dark bolometers to
be used for bias and noise monitoring.  The 2003 May chopped observations were not
deep enough to detect galaxies individually at $>3\ \sigma$ and were used only for
pointing verification by cross-correlation with the raster scan
observations.

During 2003 January, observations were made by scanning the telescope at
60\arcsec\ $\mathrm{s^{-1}}$ in right ascension and stepping in declination
by 162\arcsec\ ($\sim \frac{1}{3}$ of a field of view) between subscans (defined as a single raster scan across the sky at a fixed declination) to build up
the map.  Subsequent scans (which we define as the set of subscans needed to cover the entire declination range of the field) were taken with a $\pm 11$\arcsec\ jitter for
each 162\arcsec\ step to minimize coverage variations.  Combined with a
5$^\circ$ tilt of the bolometer array relative to azimuth and a fixed Dewar angle, such that the rotation relative to scan 
direction varies over the night, this yielded even coverage and sub-Nyquist sampling in the cross-scan
direction (declination).  Sub-Nyquist sampling was automatically achieved
in the in-scan direction (right ascension) with 50 Hz sampling of the
lock-in amplifiers.  In 2003 May, the chopped observations were made with
raster scans in azimuth and steps in elevation, but with a scan rate of
5\arcsec\ $\mathrm{s^{-1}}$ and a symmetric chopper throw of $\pm45$\arcsec\
in azimuth, with frequencies of 1 and 2 Hz.

A coverage map of the Bolocam Lockman Hole East field from 2003 January is
shown in Figure \ref{fig:coverage}, where the integration time per
10\arcsec $\times$10\arcsec\ pixel is shown.  Bolocam's 8\arcmin\
field of view is not small compared to the map size; thus, there is a large
border around the map where the coverage is reduced and nonuniform
compared to the central region.  Hence, we define a ``uniform coverage
region'' of 324 arcmin$^2$ in the center where the rms
in the integration time per pixel is 12\%.  Because rms noise varies as the
square root of the integration time, the noise dispersion is approximately
6\% in the uniform coverage region (2\% after the map has been optimally filtered as discussed in \S\ \ref{subsection:mapping}).  Our
analysis is confined to the uniform coverage region.  The observational
parameters are summarized in Table \ref{table:observationalparameters}.




\section{Data Reduction}
\label{section:datareduction}

\subsection{Basic Pipeline}

The Lockman Hole observations were reduced with a custom IDL-based software
pipe\-line.  The raw files were cleaned with a PCA sky subtraction, where an atmospheric and instrumental noise template was
generated through an eigenvector decomposition of the time stream data (\S\ \ref{subsection:cleaning}).  In the case of
the chopped observations, which are characterized by both positive and negative beams, the time streams were first demodulated followed 
by a convolution with the expected source crossing structure (first the positive beam, then the negative beam). 
This results in a positive net peak at the nominal source position with the full source amplitude and symmetric negative beams with half 
the source amplitude.

Once the cleaned time streams were obtained, a map was generated by
co-adding individual time streams, weighted by their power spectral
densities (PSDs) integrated over the spectral response to a point source.  Pointing offsets
were applied to individual observations from the global pointing model
generated from observations of submillimeter pointing sources (\S\ \ref{subsection:pointing}).  Time streams were calibrated from lock-in amplifier voltages to millijanskys using
observations of primary and secondary flux calibrators (\S\ \ref{subsection:calibration}).  The final
map was generated in right ascension and declination using sub-beam-sized
pixelization (\S\ \ref{subsection:mapping}) and Wiener filtered to maximize signal-to-noise ratio (S/N) for detections of 
point sources.

\subsection{Cleaning and Sky Subtraction}
\label{subsection:cleaning}

To facilitate removal of fluctuating atmospheric water vapor emission
(sky noise) from the bolometer signals, Bolocam was designed such that
the feed horn beams overlap maximally on the primary mirror of the
telescope and therefore sample very similar columns of atmosphere.
Thus, the sky noise, which dominates the fundamental instrument noise
by a factor of $\sim$100, is a nearly common-mode signal.  To remove
this correlated $1/f$ noise with maximum effectiveness, a PCA technique was developed.  The formalism of
the PCA analysis is standard \citep[see, e.g.,][]{murtagh87}.  Here the covariance
matrix is built from the $n$ bolometers by $m$ time elements matrix
for each subscan.  Eigenfunctions of the orthogonal decomposition that have
``large'' eigenvalues, corresponding to large contributions to the
correlated noise, are nulled and the resulting functions are
transformed back into individual bolometer time streams.  This
technique is applicable for the dim ($\lesssim$10 mJy) submillimeter
galaxies of the Lockman Hole (and other blank-field surveys) because
the source signal contributes negligibly to the sky templates and is
largely uncorrelated from bolometer to bolometer.  The PCA technique is not 
appropriate for extended sources, however, in which case the bolometers see 
correlated astrophysical signals, which are then attenuated.  The PCA
decomposition was applied to raster scan and chopped data, after chop
demodulation in the latter case.  Cosmic-ray strikes (spikes in the
time streams) are flagged and not included in constructing the
eigenfunctions.

The precise level of the cut on the large eigenvalues is somewhat
arbitrary.  The greater the number of eigenfunctions that are nulled,
the lower the resulting noise in the cleaned time stream, but the
correspondingly greater source flux density removed.  Empirically, an
iterative cut with the nulling of eigenfunctions with eigenvalues
$>3\ \sigma$ from the mean of the eigenvalue distribution produced a
balance between sky emission removal and source flux density reduction
in simulated observations by maximizing the S/N.
Because the distribution of eigenvalues for each observation is
characterized by a few outliers (typically $4-7$) at large $\sigma$-values, the overall variance of the time stream is largely dominated
by these eigenvalues, resulting in a S/N that is
insensitive to the cut threshold for $2-5$ $\sigma$.
Furthermore, the distribution of source candidates in the combined
Lockman Hole map was invariant under variations in the cut threshold
in this range.

\begin{figure}
\epsscale{1.0}
\plottwo{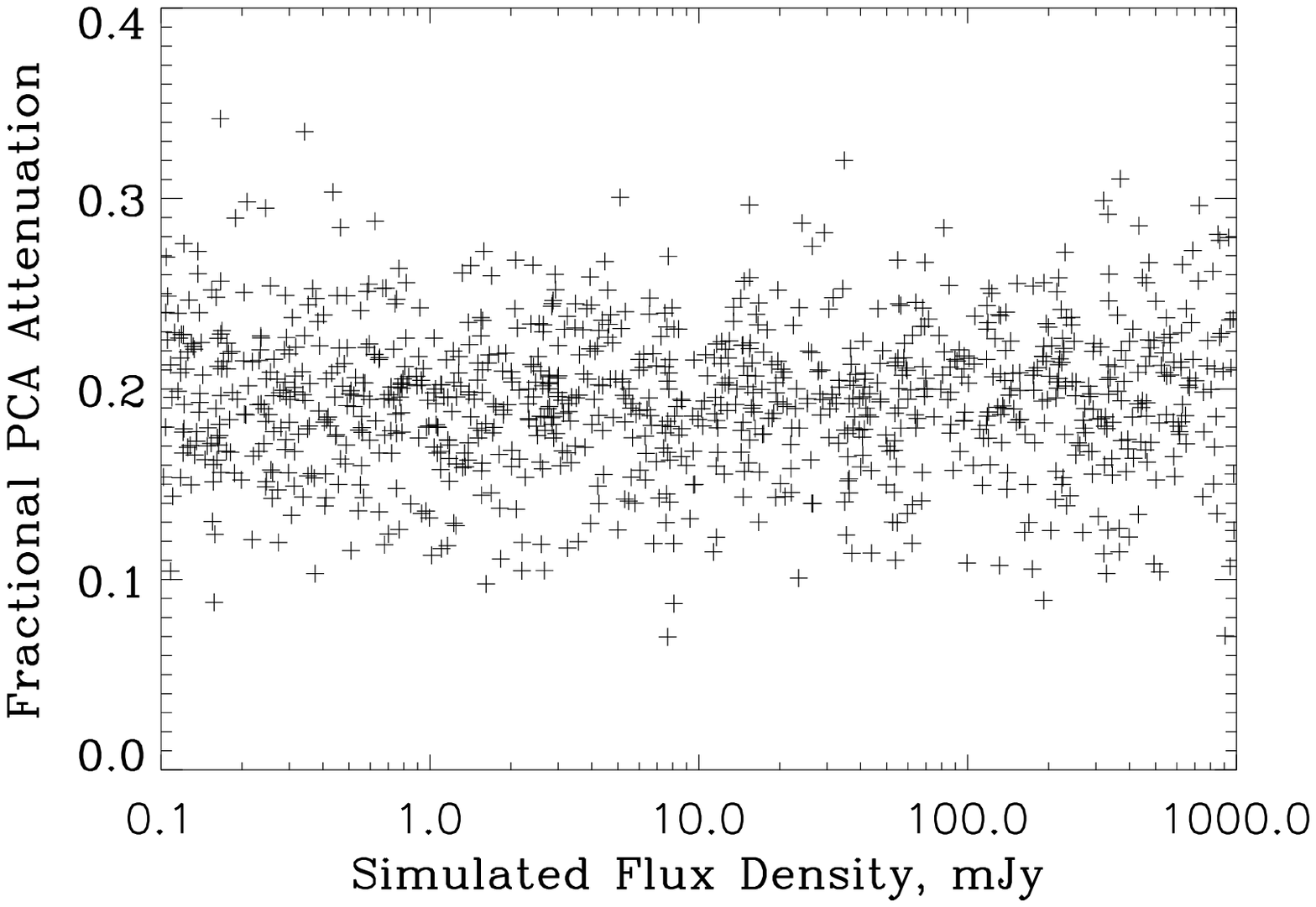}{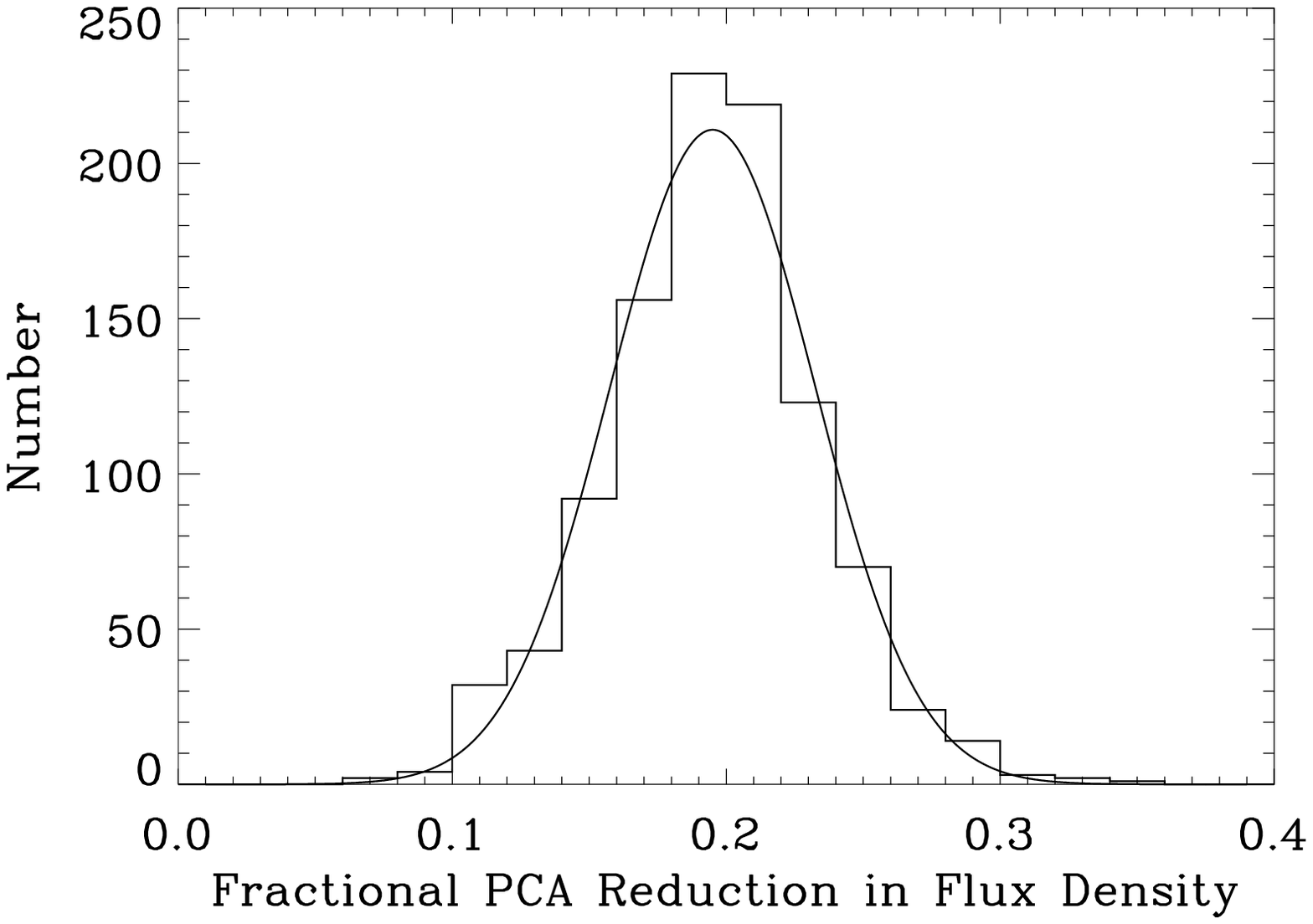}
\caption{Source attenuation by PCA cleaning as a function of injected flux density ({\it left}) and histogram with Gaussian fit ({\it right}).  Sources were injected into
  the raw time streams, which were then cleaned using the PCA.  The
  resulting source amplitudes were compared to the injected source
  amplitudes (their ratio is the fractional reduction in source flux
  density).  The attenuation of sources by PCA is 19\% with a dispersion of
  4\%, independent of flux density from 0.1 mJy to 1 Jy.}
\label{fig:fluxreduction}
\end{figure}

The PCA sky subtraction attenuates the signal from point sources in
addition to the atmospheric signal because it removes low-frequency
power from the time streams.  The amount of flux density attenuation is
determined by the number of PCA components that are removed from the
raw time streams, which is controlled by the cut on the eigenvalues: a
more aggressive cut results in greater attenuation.  Monte Carlo
simulations were done to determine the amount by which the flux
density of galaxy candidates was reduced by the cleaning.  The
simulations were done in the following manner: A fake source
(Gaussian, 31\arcsec\ FWHM) was injected into a blank Lockman Hole
map.  A simulated bolometer time stream was generated from the map of
the fake source and was added to the raw bolometer time streams of an
individual Lockman Hole observation.  The time stream data were then
cleaned with PCA and mapped in the ordinary manner.  The resulting
source was fitted by a two-dimensional Gaussian to determine the
attenuation of the injected source flux density.  This simulation was
repeated 1014 times with fake sources injected into random
observations at random positions and ranging in flux density from 0.1
to above 1000 mJy (Fig.\ \ref{fig:fluxreduction}).  The average
reduction in flux density is 0.19 with an rms dispersion of 0.04,
independent of flux density to 1 Jy.

Above 1 Jy, typical for bright pointing and flux calibrators, the amount of attenuation by PCA was found to depend on the brightness
of the fake source.  Thus, a different cleaning technique was used for
these sources.  An atmospheric noise template was generated by simply
taking an average of all $n$ bolometers for each time element.  The
mean-subtracted sky template was then correlated to each of the
individual bolometer time streams and the correlated component was
subtracted.  To prevent the correlation coefficient from being
contaminated by the calibrators, multiple scans (including
telescope turnaround time between scans) were concatenated and used together to correlate the
average sky template to each individual bolometer signal, thus
ensuring a small contribution from the point source.  A similar
analysis to that for PCA flux reduction was performed for the simple
average sky subtraction technique, yielding an average flux density
reduction of 0.07, independent of source flux, with an rms dispersion
of 0.02.

\subsection{Pointing}
\label{subsection:pointing}

Observations of planets, quasars, protostellar sources, H$_{\mathrm{II}}$ regions, and
evolved stars were used to construct separate pointing models for the
2003 January and May observing runs.  Observations of the pointing sources
were taken at the same scanning speeds as the Lockman Hole observations.
The pointing fields were generally small (scan areas of $\sim$ 4\arcmin
x4\arcmin), although several larger maps (10\arcmin x10\arcmin) were made
of Mars so that the source would pass over the entire bolometer array for
measuring relative responsivities and beam maps.  Pointing observations are generally small because source crossings are only needed in a small subset 
(15 or so) of bolometers to determine the pointing offsets.  These observations were used to map and
correct the distortion over the field of view, which is in broad agreement
with the distortion predicted by a Zemax$^{\circledR}$ ray-tracing model.  The residual rms in the raster-scanned pointing model for the ensemble of all 2003 January sources is
9.1\arcsec, although the local pointing registered to a nearby pointing source is superior.\footnote{A subsequent pointing model for a localized region of sky
yields an rms of 4.5\arcsec.}  This random pointing error results in an 18\% flux density reduction of the Lockman Hole 
galaxy candidates (analytically derived from a convolution of the 31\arcsec\ Bolocam beam with a 9.1\arcsec\ Gaussian random pointing 
error), which is corrected for in the reported flux densities (and uncertainties in these fluxes) of Table \ref{table:sourcelist}.

\begin{figure}
\epsscale{1.0}
\plotone{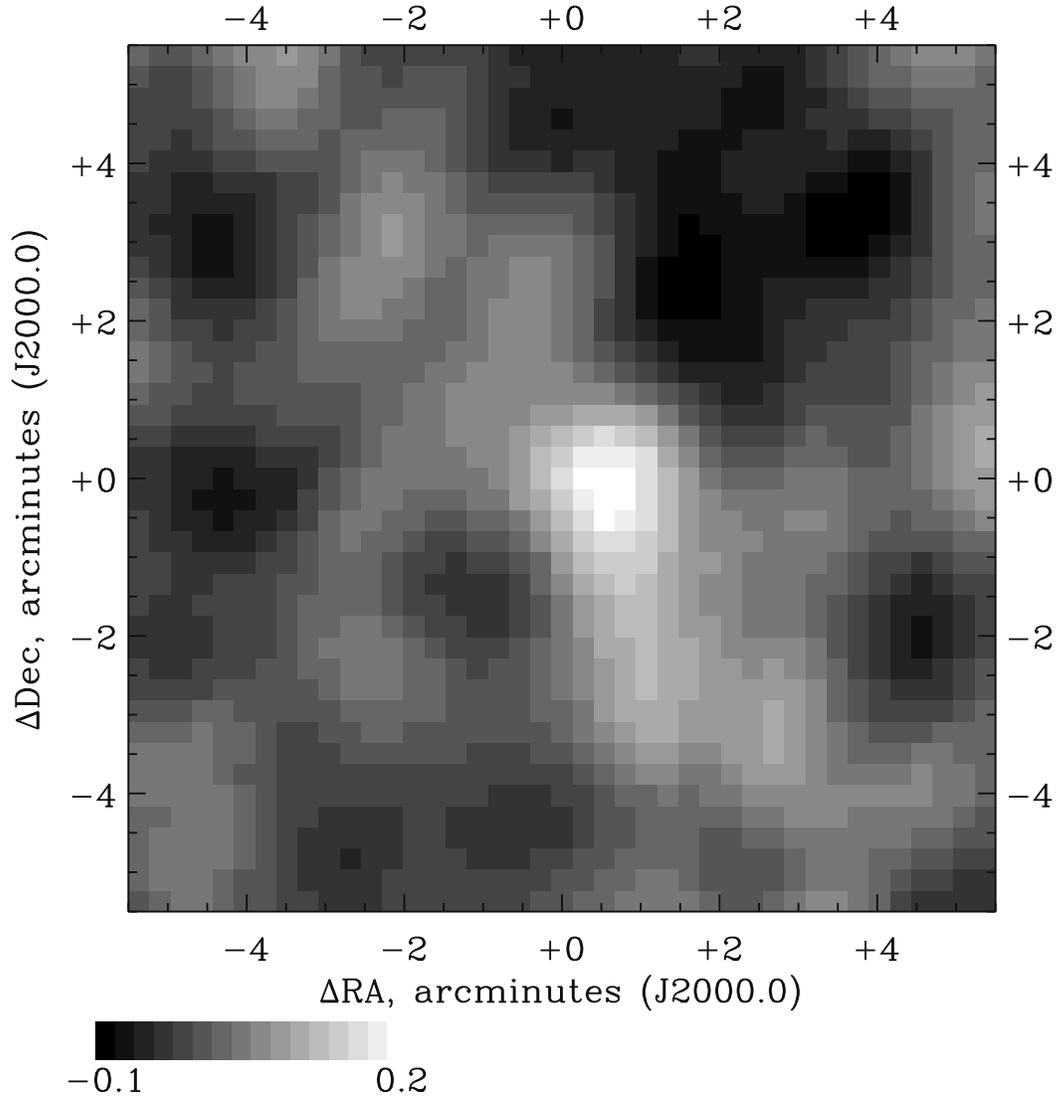}
\caption{Cross-correlation of the 2003 January and May Lockman Hole maps, pixelized at 10\arcsec\ resolution.  
While the weather was too poor during the 2003 May observations to yield any $> 3\ \sigma$ detections, the 
local pointing near the Lockman Hole was sampled substantially better than for the 
2003 January run.  The 25\arcsec\ pointing offset corresponding to the peak in the cross-correlation map was 
subsequently applied to the 2003 January data.}
\label{fig:crosscorrelation}
\end{figure}

While the 2003 January pointing observations were used to construct a pointing model that was applied to the entire sky, the region of the celestial sphere near the
Lockman Hole was not well sampled.  A pointing correction derived from sources far away ($>30^\circ$) from the Lockman Hole is therefore 
susceptible to a systematic offset.  Pointing 
observations were made much more frequently (once per hour) 
during the 2003 May run and sources near the Lockman Hole were emphasized to create an improved local
pointing model; consequently, the 2003 May pointing model near the Lockman Hole was superior to the
2003 January pointing model.  No galaxy candidates were detected at $\ge3\ \sigma$
significance in the 2003 May chopped Lockman Hole map owing to poor weather; however, it was
cross-correlated with the 2003 January map to compare the pointing models.
The cross-correlation yielded a shift of 25\arcsec\ in right ascension of
the 2003 January data with respect to the 2003 May data (Fig.\ \ref{fig:crosscorrelation}).
Because the pointing on the sky near the Lockman Hole was
substantially better and more frequently sampled for the 2003 May run, we attribute
this shift to a systematic offset in the 2003 January pointing model.  Thus, a
systematic 25\arcsec\ shift in right ascension was applied to the 2003 January Lockman
Hole map.  The need for the shift is also apparent in a comparison between
the Bolocam map and the 8 mJy SCUBA 850 $\mu$m and MAMBO 1.2 mm surveys, as several of the Bolocam galaxy
candidates become coincident with SCUBA and MAMBO sources in the overlap region of the surveys.  

Because no pointing observations were taken near the Lockman Hole, it is difficult to quantify the 
uncertainty in the 9.1\arcsec\ pointing rms.  An independent measurement of our pointing uncertainty was performed by examining both
10 VLA\footnote{The National Radio Astronomy Observatory is a facility of the National Science Foundation operated under cooperative agreement by Associated Universities, Inc.} 
radio positions coincident with Bolocam Lockman Hole galaxy candidates and the subset of 5 sources with 
additional SCUBA and/or MAMBO counterparts (see \S\ \ref{subsection:ancillary}).  The rms errors between the Bolocam and radio 
positions are $10.2^{+3.1}_{-2.4}$ and $9.3^{+4.4}_{-3.0}$ arcsec for the entire 10-source sample and 5-source subset, respectively.  The 
quoted uncertainties are the minimum length 90\% confidence intervals for 10 and 5 degrees of freedom (for both 
$\delta_{\mathrm{R.A.}}$ and $\delta_{\mathrm{decl.}}$, each of which independently determines the pointing error), respectively.  

\subsection{Flux Calibration}
\label{subsection:calibration}

Observations of primary calibrators (planets) and secondary
calibrators (protostellar sources, H$_\mathrm{II}$ regions, and evolved stars)
were used for flux calibration.  Reference planetary flux densities
were obtained from the James Clerk Maxwell Telescope (JCMT)
calibration
Web site,\footnote{See http://www.jach.hawaii.edu/jac-bin/planetflux.pl.}
and flux densities of secondary calibrators were obtained from JCMT
calibrators \citep{sandell94,jenness02}.  The flux density of IRC +10216 is
periodic; the flux density was adjusted to the epoch of observation
using the 850 $\mu$m SCUBA phase.  The reference flux densities were
corrected for the Bolocam bandpass, which is centered at 265 GHz (the
flux densities in the Bolocam band are 5\% larger than the those
quoted by the JCMT for the SCUBA 1.1 mm band).  During 2003 January,
Saturn had a semidiameter of 10\arcsec; this is not small compared to
the 31\arcsec\ Bolocam beam, so corrections for the angular extent of
Saturn were required.

\begin{figure}
\epsscale{1.0}
\plotone{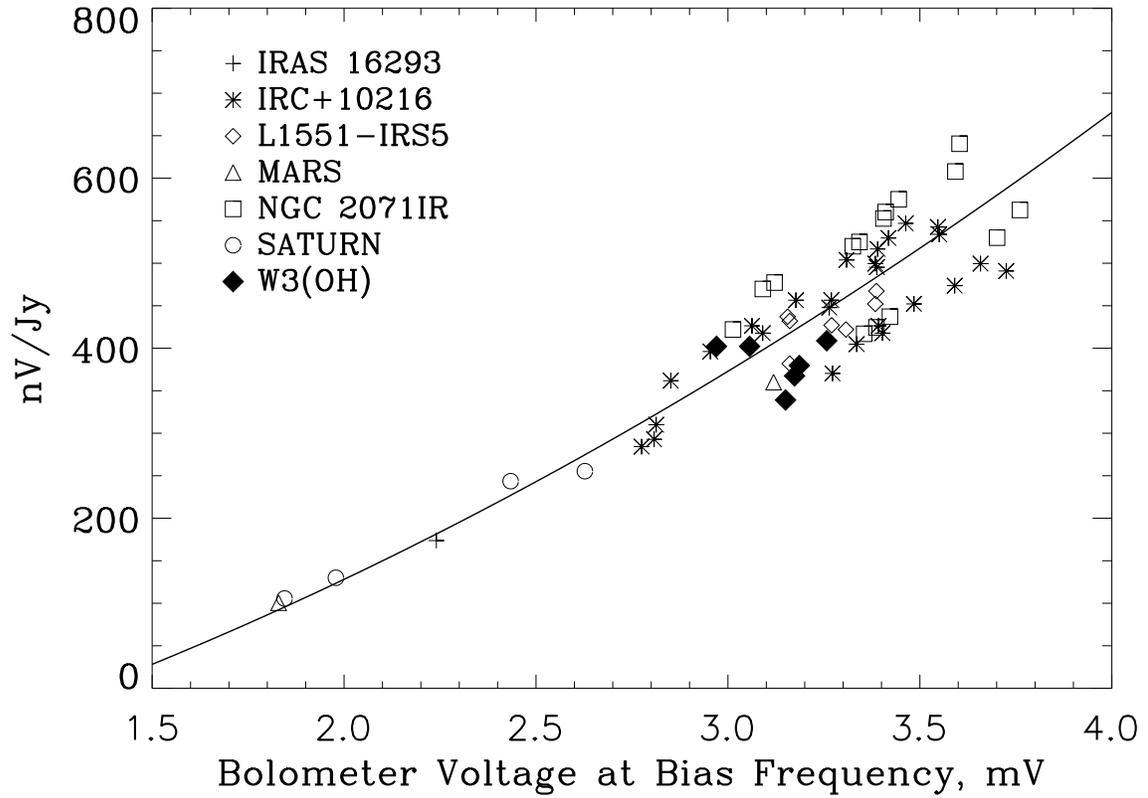}
\caption{Calibration in nV Jy$^{-1}$ (nV at the bolometer) as a function of the demodulated lock-in voltage at the AC
bias frequency, which is approximately inversely proportional to the bolometer loading.  The quadratic fit is a
minimization of the fractional error between observed and expected flux densities.  The rms of the residual dispersion 
in flux density is 9.7\%.}
\label{fig:calibration}
\end{figure}

The standard technique for flux calibration is to calibrate a given
science observation using the flux calibrator observations taken
nearest in time, which were presumably taken at similar atmospheric
opacity and air mass.  With Bolocam, we are able to use a more
sophisticated technique via continuous monitoring of the bolometer
operating resistance using the DC level of the lock-in amplifier
output signal.  The technique uses the following logic.  
The atmospheric optical loading increases as the
atmospheric optical transmission decreases, which may occur because of
changes in zenith opacity (i.e., weather) or intentional changes in
telescope elevation.  The bolometer
resistance decreases monotonically as the atmospheric optical loading
increases.  Simultaneously, the bolometer responsivity
decreases monotonically as the bolometer resistance decreases.  Thus,
the flux calibration (in nV Jy$^{-1}$, where the voltage drop
across the bolometer is proportional to its resistance), which is proportional to the product of
atmospheric transmission and bolometer responsivity, is expected to be
a monotonic function of the bolometer resistance.  This relation is
measured empirically, as shown in Figure \ref{fig:calibration}, by
plotting the flux calibration (voltage at the bolometer in nV Jy$^{-1}$
of source flux) from each of the ensemble of calibrator observations
against the median DC lock-in amplifier voltage measured during the observation.
This relation is then combined with the continuously monitored DC
lock-in signal to apply the appropriate flux calibration value during
science observations.  Note that the curve is measured only using
sources dim enough to ensure linear bolometer response; Jupiter was dropped for this reason.



%

The flux density calibration derived from Figure \ref{fig:calibration}
is biased relative to the blank-field sources by the combination of
three effects: reduction in the flux density of calibration sources
due to average cleaning, reduction in the flux density of blank-field
sources due to PCA cleaning, and reduction in the flux density of
blank-field sources due to pointing errors.
The first two effects cause the calibration curve of Figure
\ref{fig:calibration} to be shifted up by the factor
$\epsilon_{\mathrm{avg}}/\epsilon_{\mathrm{PCA}}$, where the flux reduction factors
$\epsilon_{\mathrm{avg}}$ and $\epsilon_{\mathrm{PCA}}$ are as determined
in \S\ \ref{subsection:cleaning}.  The effect of Gaussian random
pointing errors of rms $\sigma_p$ on the peak flux density of a source
observed with a Gaussian beam of width $\sigma_b$ is equivalent to a
convolution of the beam with a Gaussian of rms $\sigma_p$.  The
resulting reduction in peak height can be analytically calculated as
\begin{eqnarray}
\nonumber
\epsilon_p = \frac{\sigma_b}{\sqrt{\sigma_b^2 + \sigma_p^2}} = 0.82^{+0.09}_{-0.12}
\end{eqnarray}
for a 31\arcsec\ FWHM beam and a random pointing error of 9.1\arcsec.
The uncertainty quoted for $\epsilon_p$ is the minimum length 90\%
confidence interval obtained from the rms pointing error between the
Bolocam galaxy candidates and coincident radio sources (see \S\
\ref{subsection:ancillary}).  (While the local pointing observations
around a specific altitude and azimuth are clustered, each with a
smaller pointing error rms, the Lockman Hole observations were taken
over a large range of zenith and azimuthal angles and thus have an
overall pointing error defined by the ensemble of pointing
observations.)

Thus, the final bias in flux density is
\begin{eqnarray}
\nonumber
\epsilon = \frac{\epsilon_p \epsilon_{\mathrm{PCA}}}{\epsilon_{\mathrm{avg}}} = 0.71^{+0.08}_{-0.10}.
\end{eqnarray}
All flux densities (as well as uncertainties in these fluxes) quoted
in this paper, including the simulations of \S\
\ref{section:simulations}, have been corrected for this flux bias.
The uncertainty in the flux bias is a systematic effect that produces
a correlated shift in all source fluxes.

\subsection{Mapping and Optimal Filtering}
\label{subsection:mapping}

\begin{figure}
\epsscale{0.86}
\plotone{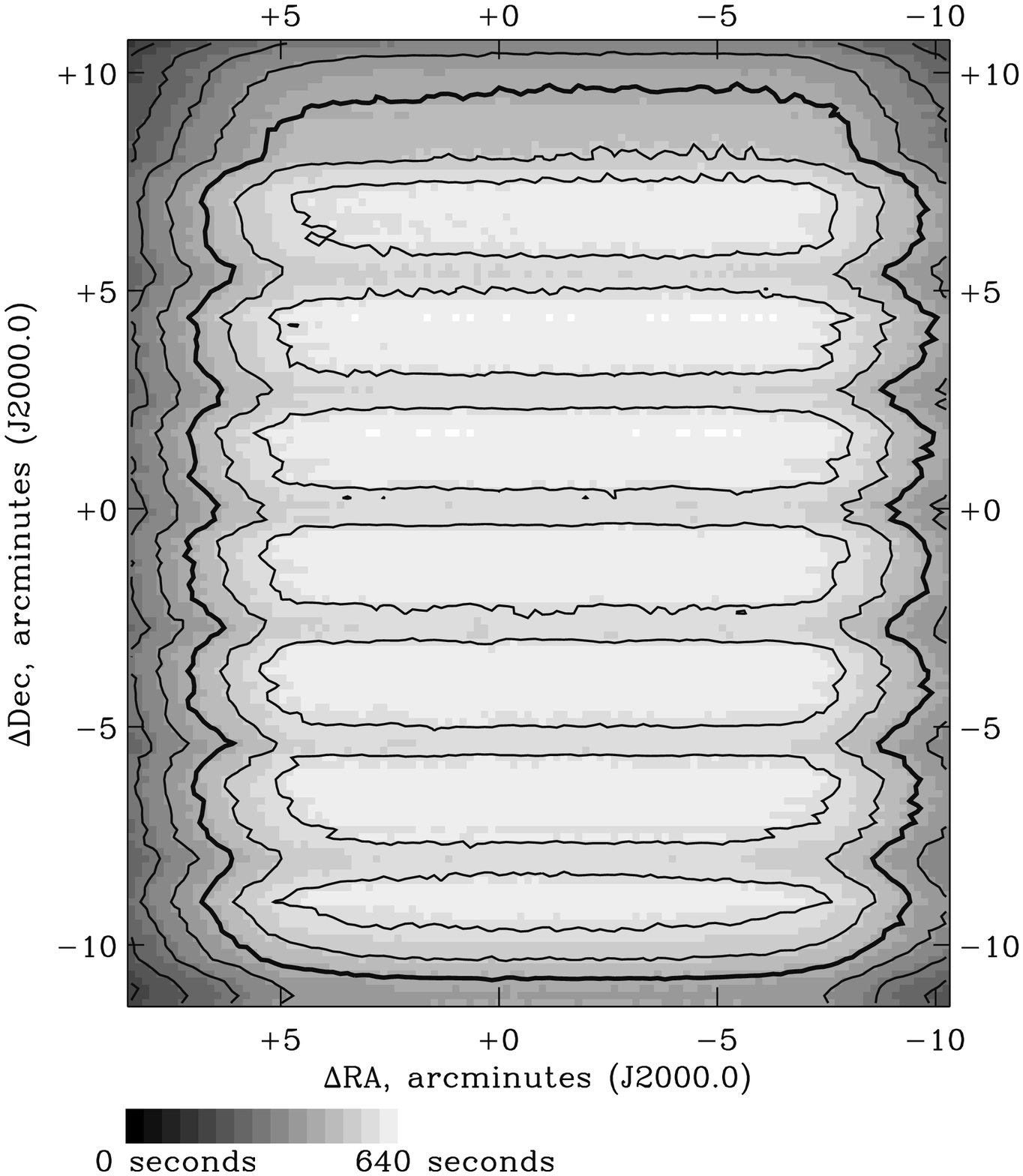}
\caption{Coverage map of the Lockman Hole East, pixelized at 10\arcsec\ resolution.  White corresponds to the
  highest level of coverage and black to the lowest level of coverage;  the
  contours are 193, 257, 322, 386, 450, 515, and 579 s
  of integration time per pixel.  The ``uniform coverage region'' corresponds
  to the thick contour at 450 s pixel$^{-1}$.}
\label{fig:coverage}
\end{figure}

Bolocam maps are built up by co-adding subscans weighted by their time
stream PSDs integrated over the spectral (temporal frequency) band of a point source at the
raster scan speed.  Data points were binned into
10\arcsec $\times$ 10\arcsec\ pixels, with approximately 30,000 hits per pixel.
Each hit represents a 20 ms integration per bolometer channel.
Four maps were created: a coverage map with the number of hits per pixel
(Fig.\ \ref{fig:coverage}); the PCA-cleaned, optimally filtered astrophysical map; a coverage-normalized map;
and a within-pixel rms map.  In the coverage-normalized map, each pixel was
multiplied by the square root of the number of hits (effectively the
integration time) in that pixel to account for the nonuniform coverage in
the map when comparing pixels.  The dispersion of the bolometer voltages (from each of the hits) within each pixel was recorded in the within-pixel rms map.  

Because the signal band of interest (point sources) does not fall throughout the entire temporal (or 
spatial) frequency range of the PSD of the data, we filter the co-added map 
with an optimal (Wiener)
filter, $g(q)$, to attenuate $1/f$ noise at low frequencies and high-frequency noise above the signal frequency:
\begin{equation}
\label{wiener}
g(q) = \frac{s^*(q)/J(q)}{\int{| s(q) |^2 / J(q) d^2q}},
\end{equation}
\noindent where $J(q)$ is the average PSD, $s(q)$ is the Fourier transform of the Bolocam 
beam shape from map space to spatial frequency ($1/x$) space, $q$, and the asterisk indicates complex conjugation.  The 
factor in the denominator is the appropriate normalization factor so that when convolved with a map, peak 
heights of beam-shaped sources are preserved.  $J(q)$ is obtained by transforming the time stream PSDs (averaged over all of the Lockman Hole observations) to a spatial PSD assuming 
azimuthal symmetry.  A two-dimensional map of 
equation (\ref{wiener}) was thus convolved with the co-added map to 
maximize S/N for detections of point sources.  An analogous filter was applied directly to the 
demodulated time streams of the chopped observations, with $s(t)$ represented by a positive and negative beam 
separated by the chop throw (90\arcsec).

\begin{figure}
\epsscale{0.78}
\plotone{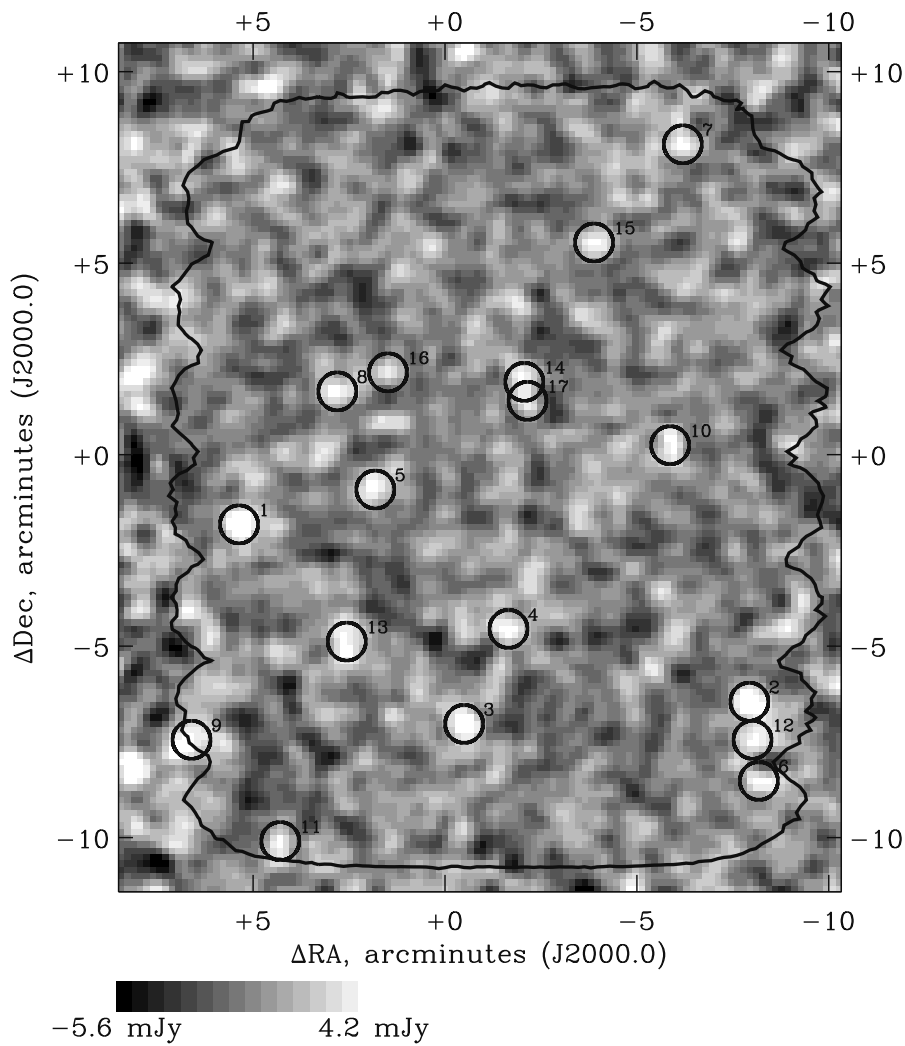}
\caption{Bolocam map of the Lockman Hole East.  The field is centered on
  $\mathrm{R.A.} = 10^\mathrm{h}52^\mathrm{m}08^\mathrm{s}.82$, $\mathrm{decl.} = +57^\circ21'33\arcsec.80$ (J2000.0).  The map pixels are
  10\arcsec $\times$10\arcsec\, and the map rms is 1.4 mJy.  The
  uniform, high-coverage region of the map is the inner 324 arcmin$^2$.  This map has been cleaned and optimally filtered for
  point sources.  The 17
  Bolocam sources detected at $>3\ \sigma$ are indicated by thick circles.
  The bright spot at +8', -8' is not listed as a detection because it falls outside of the uniform coverage region ({\it black 
  contour}).}
\label{fig:map}
\end{figure}

The cleaned, co-added, optimally filtered map is presented in Figure \ref{fig:map}.
There is a perimeter a few arcminutes wide around the map that does not lie
within the uniform coverage region (cf.\ Fig.\ \ref{fig:coverage}).  There are 17 galaxy candidates at $>3\ \sigma$,
apparent as unresolved bright spots, numbered in order of
decreasing brightness.  Six false detections are expected from simulations (discussed in detail in \S\ \ref{section:simulations}).  There are no negative sidelobes associated with the
source candidates because the observations are not chopped.  The 850 $\mu$m SCUBA 8
mJy \citep{scott02} and 1.2 mm MAMBO \citep{greve04} surveys cover patches with radii of $\sim5$\arcmin\ and $\sim7$\arcmin\ in the center of
the map (central 122 and 197 arcmin$^2$, respectively).  A comparison of the maps is given in \S\ \ref{section:sourcelist}.

\section{Source List}
\label{section:sourcelist}

\subsection{Source Extraction}
\label{subsection:extraction}

Source extraction was performed on the PCA-cleaned, optimally filtered,
coverage-normalized map consisting of all the raster scan observations
co-added together.  The algorithm was begun by doing a cut on the uniform
coverage region, defined as the set of those pixels for which (1) the
coverage is $\ge70$\% of the maximum per pixel coverage and (2) the within-pixel rms is less than 2 $\sigma$ from the mean within-pixel rms.
The uniform coverage region is a contiguous region in the center of the map.

Next, an rms in sensitivity units (the flux density of
each pixel times the square root of the integration time for that pixel in units of mJy s$^{1/2}$) was computed in the 
uniform coverage region.  This rms is valid
for the entire uniform coverage region since variations in coverage have been
accounted for by the $t_i^{1/2}$ coverage normalization, where $t_i$ is the total integration time for pixel $i$.  All 
pixels with coverage-normalized flux densities exceeding $3\ \sigma$ (``hot pixels'')
were flagged as potential sources.  Then hot pixels were grouped
into multi-pixel sources by making the maximal group of adjacent hot pixels, including those within
$\sqrt{2}$ pixels (i.e., diagonally adjacent).  The right ascension and
declination of the source candidates were computed
by centroiding two-dimensional Gaussians on the groups.  Because convolution of the map with the Wiener filter properly weights 
the flux contribution from each pixel, the best estimate of the source flux density in the optimally filtered map 
is given by the peak value in the group.

A histogram of the pixel values in the uniform coverage region is shown in Figure \ref{fig:histogram}.
The quantity that is plotted is the pixel sensitivity, with the scaling by $t_i^{1/2}$ accounting for the
nonuniform coverage in the map.  Note that the sensitivity histogram should not be interpreted as instrument 
sensitivity as the histogram uses an optimally filtered (smoothed) map but scales by sub-beam-sized integration 
times.  The negative side of the histogram, plotted logarithmically, is extremely Gaussian.  A Gaussian fit to the Bolocam noise-only (jackknife) distribution
is overplotted by a solid line (see \S\ \ref{subsection:nulltests}), indicating a clear excess on the positive
side with respect to the Gaussian.  The galaxy candidates make up this
excess.  Since the pixels are 10\arcsec $\times$10\arcsec\ in size and the
beam size is 0.30 arcmin$^2$, there are approximately 11 pixels per
source candidate.

The source candidate list is presented in Table \ref{table:sourcelist}, where the flux densities
are listed in order of decreasing brightness in the fifth column.
Seventeen galaxy candidates were detected at $>3\ \sigma$ significance, with the
brightest being 6.8 mJy.  Seven of the candidates were detected at
$>3.5\ \sigma$ significance.  The flux densities of the source candidates
were attenuated by the PCA cleaning; their corrected flux densities are
listed (see \S\ \ref{subsection:cleaning}).  The source candidate list is compared
to the 850 $\mu$m SCUBA 8 mJy and 1.1 mm MAMBO surveys in \S\ \ref{subsection:ancillary}.

\subsection{Tests for Robustness of Galaxy Candidates}
\label{subsection:nulltests}

\begin{figure}
\epsscale{1.0}
\plotone{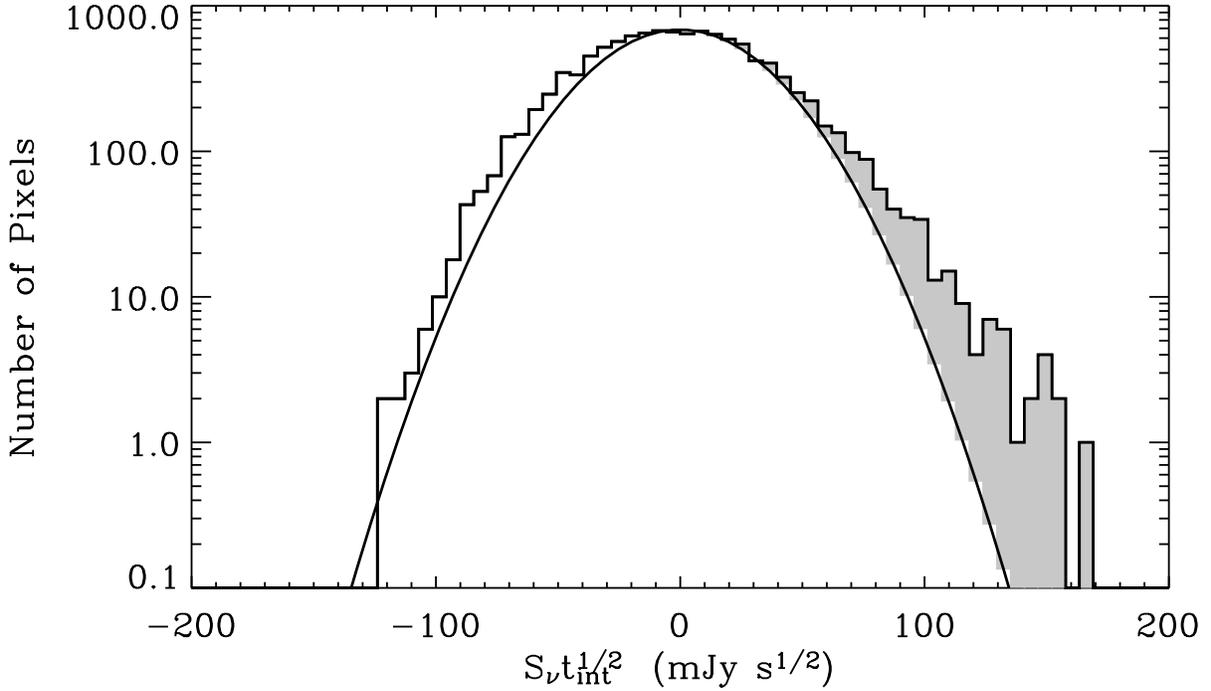}
\caption{Histogram of the pixel sensitivities, defined as the pixel flux density,
  $S_\nu$, times the integration time per pixel, $t^{1/2}_{int}$.  The solid line is a Gaussian
  fit to the jackknifed histogram of Fig.\ \ref{fig:jackknife}.  The shaded area indicates the emission due to galaxy candidates in excess of that expected from map 
  noise.  The negative side of the histogram is slightly broader than the fit to the jackknifed histogram owing to the presence of the galaxies (confusion noise) on both the positive and 
  negative side.  
  The small negative offset of the peak of the distribution is expected, as the mean of the entire map (and therefore the histogram) is constrained to have mean zero from both
  the high-pass filter in the Bolocam electronics and sky subtraction. }
\label{fig:histogram}
\end{figure}

\begin{figure}
\epsscale{1.0}
\plotone{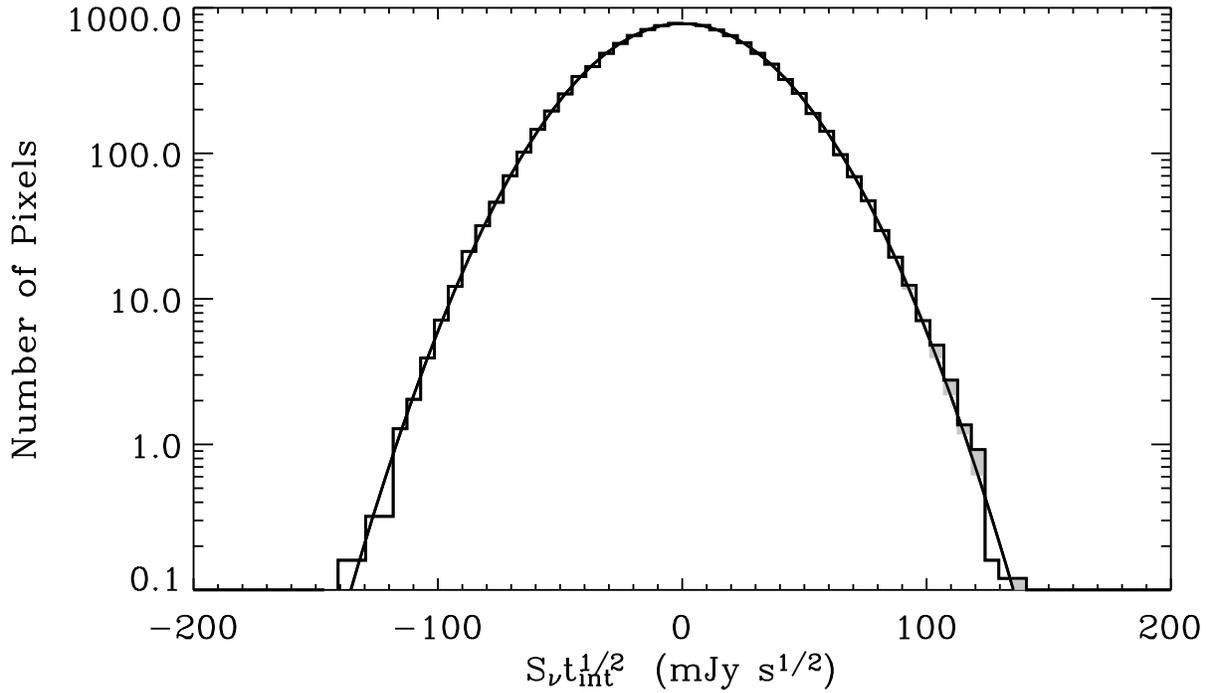}
\caption{Jackknife histogram.  In the jackknife test, 50\% of the
  observations were randomly chosen and co-added together, while the remaining
  50\% of the observations were co-added into a second map.  The two maps
  were then differenced.  This was repeated 21 times with the observations
  randomly selected independently each time, and the histograms were
  averaged.  The thick solid line corresponds to a Gaussian fit to the jackknifed histogram.  The shaded region indicates
  the positive excess, which is insignificant.}
\label{fig:jackknife}
\end{figure}

Two tests were carried out to check the robustness of the galaxy
candidates.  The first test was a jackknife test in which 50\% of the
observations were randomly chosen and co-added together into a map and the
remaining 50\% of the observations were co-added into a second map.  If the
source candidates are real and coherent over multiple observations, then
the positive-side excess of the histogram in Figure \ref{fig:histogram} should disappear when the two maps are differenced.  
Conversely, if the source candidates arise from spurious events in individual observations, such as cosmic-ray 
strikes, then the excess would not disappear when the two maps are differenced.  This algorithm was repeated
21 times with the first 50\% of the observations randomly selected
independently each time, and the histograms were averaged.  For such an algorithm, one expects the noise realizations to be 
approximately independent; the actual correlation was measured to be $\sim$4\%.  The result is
shown in Figure \ref{fig:jackknife}.  A Gaussian distribution fits the jackknife histogram extremely well.  The absence of a positive-side excess indicates that the source candidates in the Wiener-filtered map are 
common to all observations.  The negative side of the real map histogram (cf.\ Fig.\ \ref{fig:histogram}) is slightly broader because confusion noise from sources below our threshold is 
absent in the jackknife 
histogram.  Similar histograms result 
from jackknife tests of scan direction (+R.A.\ vs. -R.A.), intranight variations (cuts on local 
sidereal time), and night-to-night variations, indicating that the galaxy candidates are not caused by 
systematic effects, such as scan-synchronous or elevation-dependent noise.  This strong statistical test indicates that 
the Bolocam source candidates are real.

\begin{figure}
\epsscale{1.0}
\plottwo{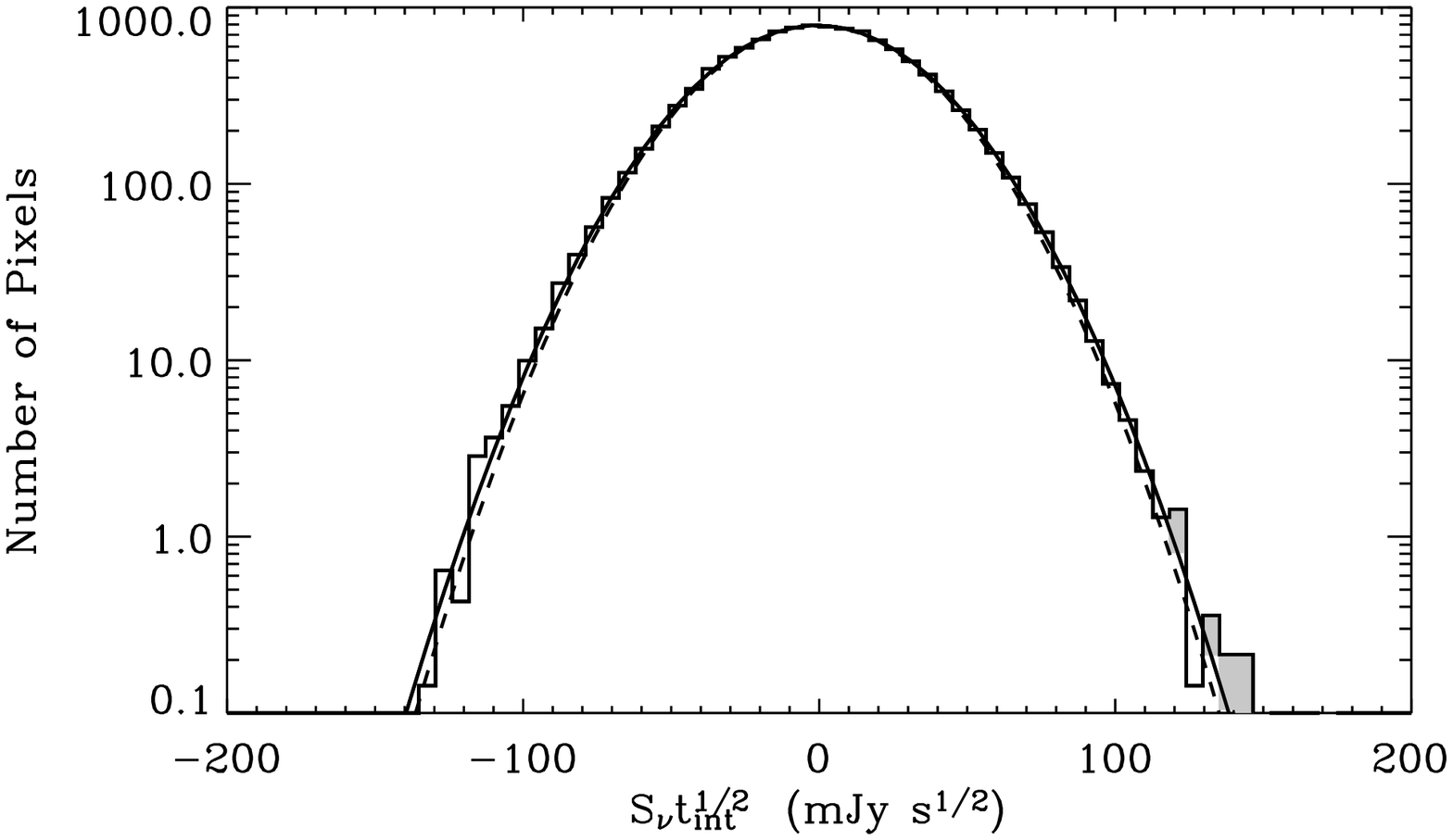}{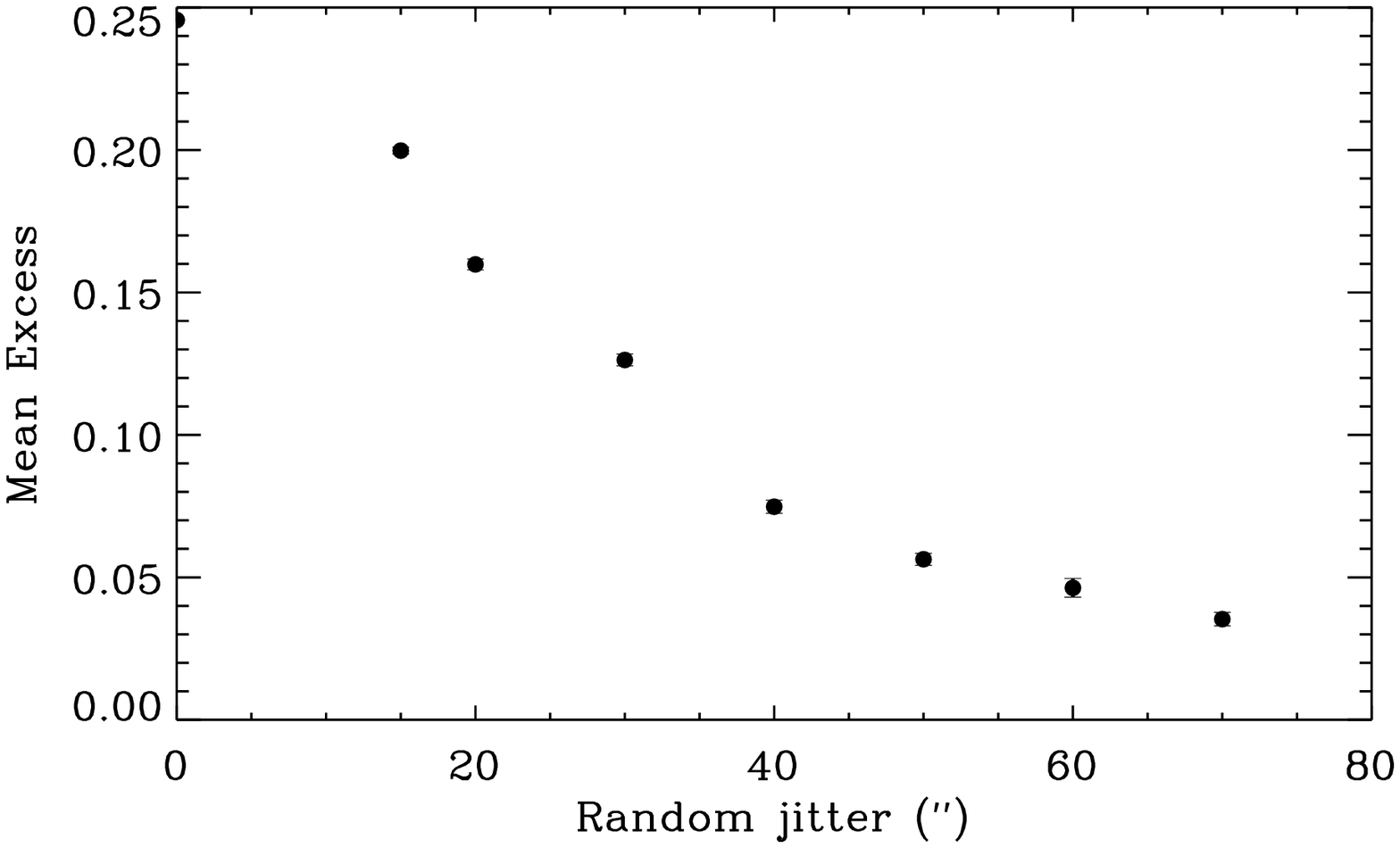}
\caption{{\it Left}: Pointing-jittered histogram.  In the pointing jitter test,
  individual maps were made from each observation,
  then co-added with 60\arcsec\ amplitude offsets with random directions (phases).  The solid line is a Gaussian fit
  to the negative side of the histogram and mirror imaged to the right side
  of the histogram, indicating that the galaxy candidates have
  disappeared.  The dashed line corresponds to a Gaussian fit to the jackknifed histogram of Fig.\ \ref{fig:jackknife}.  {\it Right}:  Decreasing positive-side excess as a function of
  random jitter amplitude.  In this case, the amplitude of the jitter was varied from 15\arcsec\ to 70\arcsec.  The excess is
  defined as the fractional increase in the rms of the jittered histogram compared to the rms of the jackknife 
  distribution of Fig.\ \ref{fig:jackknife}.  Sixteen iterations were performed at each
  jitter amplitude with the average plotted in the figure.  The statistical uncertainty in each mean excess is smaller than the size of the plotted point.}
\label{fig:jitter}
\end{figure}

A second test was performed to verify that the source candidates arise from
the co-addition of many observations rather than from spurious events.  In this
test, individual maps were made from each of the 259 observations.  These maps
were then co-added with fixed-amplitude offsets with random directions
(phases).  The expectation of this null test is that sources coherent over multiple observations are smeared 
out onto rings of a fixed radius, resulting in the disappearance of the positive-side excess.  (The 
positive-side excess will be distributed over many pixels and therefore spread over many bins of the histogram.) 
 Source candidates arising from isolated spurious events or characterized by length scales 
much larger than the Bolocam beam will merely be moved or negligibly broadened, leaving the histogram unchanged.  
Sixteen iterations were performed at each jitter amplitude ranging from 15\arcsec\ to
70\arcsec\ (Fig.\ \ref{fig:jitter}).  The rms of the jittered histograms in excess of the rms of the 
jackknife distribution of Figure \ref{fig:jackknife} continues to decline out to a
random jitter of 70\arcsec.  The excess does not go to zero at large
amplitudes because the sources are spread out onto annuli with finite
radii and will still be present at a low level.  Since the area of the annulus increases as $r$, the excess 
should drop as $r^{-1}$ (at large jitter amplitudes where the beams do not overlap), as indicated by Figure \ref{fig:jitter}.  This null test confirms that the excess 
variance (the positive-side excess in the histogram of Fig.\ \ref{fig:histogram} from source candidates) is contributed to by the ensemble of observations and has a small characteristic 
length scale (corresponding to point sources).

\subsection{Comparison With Other Submillimeter and Millimeter-Wave Surveys}
\label{subsection:ancillary}

The Bolocam galaxy survey provides a unique contribution to the current state of submillimeter galaxy surveys.  The 850 $\mu$m JCMT SCUBA 8 mJy survey (Scott et al.\ 2002; hereafter SCUBA 
survey), 
with a 14\arcsec\ 
beam, implemented a jiggle map strategy with a 30\arcsec\ chop throw over 122 arcmin$^2$ to an rms of 2.5 mJy beam$^{-1}$.  
The 1.2 mm IRAM MAMBO survey (Greve et al.\ 2004; hereafter MAMBO survey), with a 10.7\arcsec\ beam, scanned at 5\arcsec\ s$^{-1}$ with a chop throw of 36\arcsec$-$45\arcsec\ and a 
chop 
frequency of 2 Hz over 197 arcmin$^2$ to an rms of 0.6$-$1.5 mJy 
beam$^{-1}$.  
Bolocam's 60\arcsec\ s$^{-1}$ raster scan 
strategy (without chopping) facilitated a large 324 arcmin$^2$ survey to a uniform rms of 1.4 mJy beam$^{-1}$ (Wiener filtered for detection of point sources).  Using a model SED based on 
nearby, dusty, star-forming 
galaxies 
(see \S\ \ref{subsection:luminosities}) gives relative flux densities of 1\,:\,2.0\,:\,0.8 and relative rms of 1\,:\,0.9\,:\,0.6$-$1.4 for the Bolocam, SCUBA, and MAMBO surveys, 
respectively, for a galaxy redshift of 
$z=2.4$ (with the range given for MAMBO due to 
nonuniform noise).  

\begin{figure}
\epsscale{0.94}
\plotone{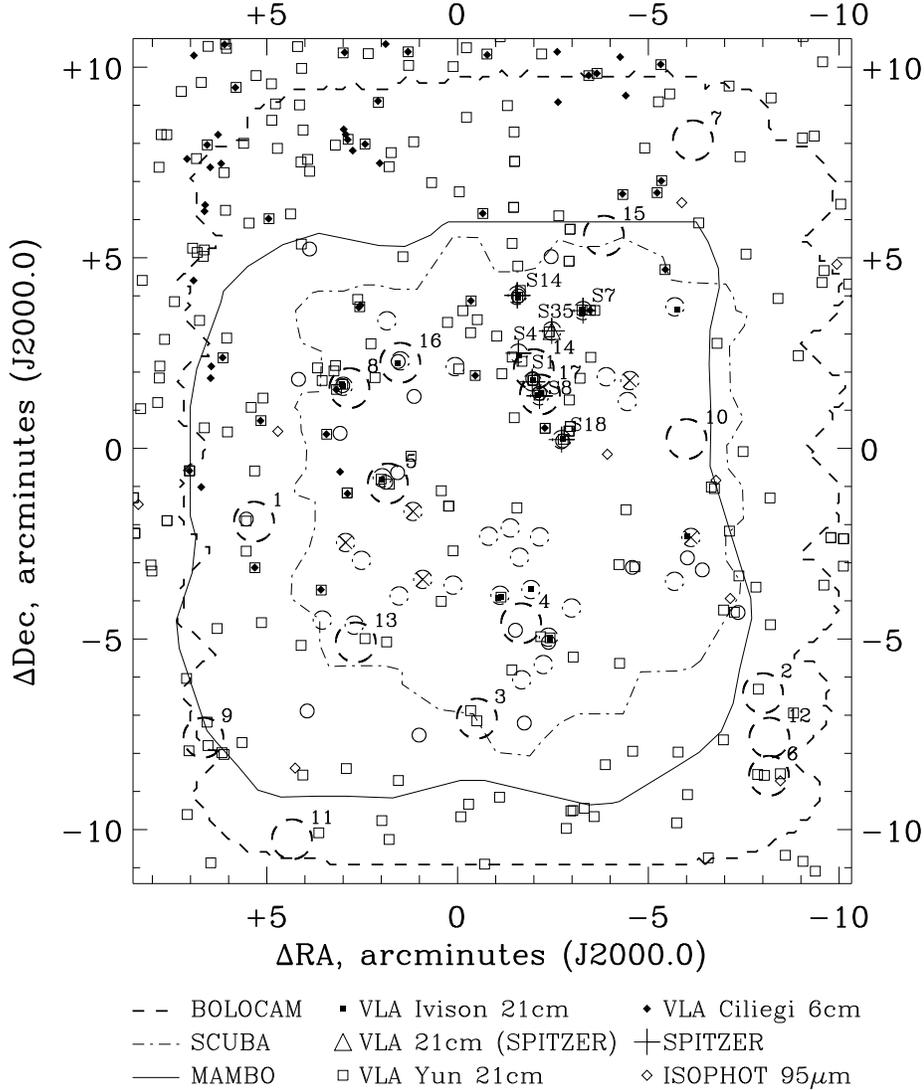}
\caption{Galaxy surveys in the Lockman Hole East region.  The 17 Bolocam submillimeter galaxy candidates have been labeled in order of brightness.  The circle radii of 
Bolocam, SCUBA \citep{scott02}, and MAMBO \citep{greve04}
detections correspond to twice the beam sizes and stated pointing
errors added in quadrature to indicate an approximate region of
astrometric uncertainty and source confusion.
The good coverage regions of Bolocam, SCUBA and MAMBO are shown.  VLA radio sources
of \cite{ciliegi03}, M.\ Yun (2004, private communication), and \cite{ivison02} are identified by filled diamonds, open squares, and filled squares, respectively.  The five 
crossed-out SCUBA 
sources are those retracted by \cite{ivison02}.  Open diamonds correspond to 95 $\mu$m ISOPHOT detections of \cite{rodighiero04}.  Plus signs correspond to SCUBA 
sources 
detected by {\it Spitzer} IRAC and/or MIPS observations \citep{egami04}.  Reexamination of the 20 cm \cite{ivison02} VLA radio map by the {\it Spitzer} group 
\citep{egami04} reveals a source ({\it triangle}) coincident with SCUBA source LE850.35.}
\label{fig:ancillary}
\end{figure}

Figure \ref{fig:ancillary} provides a cumulative overview of recent far-infrared, submillimeter, and radio observations of the Lockman Hole.  The circles of Bolocam, SCUBA, and 
MAMBO observations correspond to 2 $\sigma$ confidence regions of position, including both beam sizes and stated pointing errors.  The 6 cm VLA radio sources 
of \cite{ciliegi03} and unpublished 21 cm VLA sources of M.\ Yun (2004, private communication), with average noise levels of 11 and 10$-$15 $\mu$Jy beam$^{-1}$, 
respectively, are 
identified.  The M.\ Yun (2004, private communication) radio field covers the entire Bolocam good coverage region, while the center of \cite{ciliegi03} 
observations is offset to the 
northwest, with an overlap of 
approximately 130 arcmin$^2$.
Also shown are the 20 published radio sources from deep 21 cm VLA observations (average noise level of 4.8 $\mu$Jy beam$^{-1}$) from \cite{ivison02} that are 
coincident with SCUBA sources, as well as an additional 21 cm VLA source discovered by \cite{egami04} from the reexamination of the \cite{ivison02} map.  Infrared 
detections from a 
recent wide-field 95 $\mu$m ISOPHOT survey \citep{rodighiero04} and recent {\it Spitzer} detections of SCUBA sources \citep{egami04} are also identified.  Five SCUBA 
sources from the \cite{scott02} catalog (LE850.9, 10, 11, 15, 20) were retracted by \cite{ivison02} on the basis of large $\sigma_{\mathrm{850\mu m}}$ values (and lack of 
radio 
identifications) and are depicted by crosses through them.  

Examination of Figure \ref{fig:ancillary} shows discrepancies in detections between the surveys.  Table \ref{table:ancillary} summarizes 
the coincident detections between Bolocam 1.1 mm, SCUBA 850 $\mu$m, 
MAMBO 1.2 mm, and VLA radio observations.  Each row in the table corresponds to the fractional number of counterparts detected by each 
survey.  The five SCUBA sources retracted by \cite{ivison02} are not included in this comparison.  The coverage of each survey was taken into account, with only 
the overlapping uniform coverage regions considered.  The surveys have a wide range of agreement, ranging from 23\% (7 of 31 SCUBA 
sources detected by Bolocam) to 75\% (6 of 8 Bolocam sources detected by SCUBA).  Six of the 17 Bolocam detections are galaxies previously 
detected by the SCUBA 8 mJy survey.  Of the remaining 11 Bolocam sources, 9 of them lie outside the SCUBA 8 mJy survey region.  Similarly, 7 of the 11 Bolocam 
sources present within the MAMBO good coverage region were detected with MAMBO at 1.2 mm.  Two of the 4 Bolocam source candidates not detected by MAMBO have expected 1.2 mm flux densities (from the model SED of 
\S\ \ref{subsection:luminosities}) below the MAMBO detection threshold for $z = 2.4$.  The large fraction of Bolocam sources detected by SCUBA and MAMBO suggests that these submillimeter 
galaxy candidates are real.  
The impact of the converse of this statement is less clear: The majority of SCUBA and MAMBO sources were not detected by Bolocam, although 17 out of 24 nondetected SCUBA sources and 7 out of 15 nondetected MAMBO 
sources have expected 1.1 mm flux densities (from the aforementioned model SED) below the Bolocam detection threshold, nor is there a strong correlation between SCUBA and MAMBO sources.  Some of these sources may not be 
real or may not be modeled well by the assumed SED.  Furthermore, the majority (65\%) of Bolocam source candidates have at least one radio coincidence 
(Ivison et al.\ 2002; Ciliegi et al.\ 2003; M.\ Yun 2004, private communication), although a 34\% accidental detection rate is expected.  (This accidental detection rate is 
the Poisson likelihood that one or more of these known radio sources, randomly distributed, fall within the 2 $\sigma$ confidence region 
of the Bolocam beam.)$\:$  To help verify the 9.1\arcsec\ pointing rms of \S\ \ref{subsection:pointing}, the rms positional 
error of the Bolocam galaxy candidates compared to the VLA radio positions was calculated for both all Bolocam sources with radio 
counterparts and the subset of sources (1, 5, 8, 16, and 17) with additional SCUBA and/or MAMBO detections.  (Bolocam galaxy 
candidate 14 was excluded owing to source confusion.)

Five of the Bolocam source candidates show no counterparts in the other surveys.  These may be false detections, although four of these 
candidates are near the edge or outside of the SCUBA and MAMBO good coverage regions, which may explain the lack of additional submillimeter detections.  It is possible that one or more of 
these four Bolocam 
source candidates without radio counterparts may instead be 
sources at high redshift ($z > 3$), where the positive {\it K}-correction (sharp drop in flux density with 
increasing redshift) causes dim radio counterparts.  Four SCUBA detections that 
have at least two detections from MAMBO, {\it Spitzer}, and VLA were not detected by Bolocam, although the corresponding pixels in the Bolocam map have flux values just below 
the 4.2 mJy detection threshold for two of these nondetections.  A description of each Bolocam detection (as well as nondetections) follows in the next section.  A follow-up paper is in preparation that 
will include a more detailed discussion on individual sources, including redshift/temperature constraints.

The Bolocam 1.1 mm beam solid angle is 0.30 arcmin$^2$ and
the uniform coverage region of the map is 324 arcmin$^2$.  There are thus approximately 1000 beams in the 
map.   With 17 source candidates, or $\sim50$ beams per source, source confusion is not a serious issue.  We define ``source confusion'' here as a high spatial density of detected sources 
that 
makes it difficult to distinguish individual sources.  This should not be 
confused with ``confusion noise'' from sources below the detection threshold (discussed in detail in \S\ \ref{subsec:ConfusionNoise}).  Nevertheless, source confusion exists at some 
level because several Bolocam sources are either closely spaced or near multiple SCUBA detections.  While the clustering properties of submillimeter galaxies remain uncertain, there exists 
tentative evidence from both two-dimensional angular correlation functions \citep{greve04, scott02, borys03} and clustering analyzed with 
spectroscopic redshift distributions \citep{blain04} that suggests strong clustering with large correlation lengths \citep[as well as correlation to other classes of high-redshift galaxies, 
including Lyman break galaxies and X-ray loud AGNs;][]{almaini03}.  We do not attempt to quantify source confusion here but address it in a paper in preparation.


There are 24 8 mJy SCUBA 850 $\mu$m and 15 MAMBO 1.2 mm sources within our survey region that we did not detect. 
Statistically, however, we detect the aggregate average of these at significances of 3.3 and 4.0 $\sigma$, 
respectively, at $\lambda=1.1$ mm.  This was done by measuring the distribution of ``sensitivity" values (scaled by 
$t_i^{1/2}$) for the Wiener-filtered map pixels that coincide with SCUBA or MAMBO
sources for which we found no excursion above our detection threshold.
If no subthreshold ``counterparts" are present in these pixels, the
sensitivity values should follow the noise distribution of the map
(albeit truncated at 3 $\sigma$).  Such a distribution has a mean value of -0.004 $\sigma$.
In the data, we find that the sensitivity values for these map pixels
have mean values of $1.0 \pm 0.3$ and $0.8 \pm 0.2\ \sigma$ for the SCUBA and MAMBO
nondetections, respectively.  Assuming that uncertainties are Gaussian
distributed, the probabilities of such large nonzero means to have
arisen from pure noise are very low, $1.7 \times 10^{-4}$ and $2.5 \times 10^{-4}$, respectively.  Thus,
we have statistically detected the ensemble of SCUBA and MAMBO sources below our
threshold at the 3.3 and 4.0 $\sigma$ confidence levels, respectively.



\subsection{Bolocam Detections}

The Bolocam detections are as follows:

\noindent {\it Bolocam.LE1100.1.---}This 6.8 mJy galaxy candidate has Bolocam and MAMBO detections but falls outside of region covered 
by SCUBA observations.  A strong 20 cm VLA radio observation (M.\ Yun 2004, private communication) exists within both the Bolocam and MAMBO confidence 
regions.

\noindent {\it Bolocam.LE1100.2.---}This 6.5 mJy Bolocam detection has a coincident 20 cm radio (M.\ Yun 2004, private communication) detection.  The source 
lies outside good coverage regions of the SCUBA, MAMBO, and \cite{ciliegi03} VLA observations.

\noindent {\it Bolocam.LE1100.3.---}This source has a 6.0 mJy Bolocam detection with two 20 cm radio (M.\ Yun 2004, private communication) plausible 
counterparts but lacks 
a MAMBO detection.  The source lies outside the good coverage regions of the SCUBA and \cite{ciliegi03} VLA observations.

\noindent {\it Bolocam.LE1100.4.---}This 5.2 mJy galaxy candidate has Bolocam and MAMBO detections with no SCUBA or radio detections.  Several SCUBA sources (with radio counterparts) are in 
close 
proximity to this source; however, the coincident MAMBO source (with a comparatively small 10.7" beam FWHM) confirms the Bolocam detection at this location.

\noindent {\it Bolocam.LE1100.5.---}This 5.1 mJy galaxy candidate has Bolocam, SCUBA, and VLA \citep{ivison02} detections with two coincident MAMBO and M.\ 
Yun (2004, private communication) 
detections.  

\noindent {\it Bolocam.LE1100.6.---}Bolocam detection 6 (5.0 mJy) has three potential 20 cm VLA radio counterparts (M.\ Yun 2004, private communication), and 
a 95 
$\mu$m ISOPHOT \citep{rodighiero04} detection within the Bolocam confidence 
region.  No SCUBA or MAMBO coverage (or Ciliegi et al.\ [2003] VLA coverage) exists for this detection.

\noindent {\it Bolocam.LE1100.7, 10, 11, 12, 15.---}These five sources (4.9, 4.7, 4.6, 4.6, and 4.4 mJy, respectively) are Bolocam galaxy candidates with no other coincident detections 
(including radio counterparts), although sources 7, 11, 12, and 15 fall near the edge or outside of the
SCUBA, MAMBO, and VLA \citep{ciliegi03} good coverage regions.  These sources may be false detections, since six are statistically 
expected from simulations (see \S\ \ref{section:simulations}), or possibly galaxies at high redshift ($z > 3$).  The likelihood of sources 10, 11, 12, 
and 15 being false detections is enhanced by the 
fact that their flux densities are near the 4.2 mJy threshold.    

\noindent {\it Bolocam.LE1100.8, 16.---}These submillimeter galaxies have Bolocam (4.8 and 4.1 mJy, 
respectively), SCUBA, MAMBO, and VLA (8, 16, Ivison et al.\ 2002; 8, Ciliegi et al.\ 2003; 8, M.\ Yun 2004, private communication) 
detections.

\noindent {\it Bolocam.LE1100.9.---}This galaxy candidate has a 4.8 mJy Bolocam detection with three 20 cm radio detections (M.\ Yun 2004, private 
communication) 
within the Bolocam confidence region.  There is no SCUBA, MAMBO, or \cite{ciliegi03} VLA coverage in this region.

\noindent {\it Bolocam.LE1100.13.---}This 4.5 mJy Bolocam detection has SCUBA and VLA radio (M.\ Yun 2004, private communication) counterparts, but no 
\cite{ivison02} or \cite{ciliegi03} VLA detections. 

\noindent {\it Bolocam.LE1100.14, 17.---}These two closely spaced Bolocam detections (4.4 and 4.0 mJy, respectively) have 
numerous other detections, including three SCUBA sources, two MAMBO sources, and multiple VLA radio sources (S1, S4, S8, Ivison et al.\ 2002; S1, S4, S8, 
M.\ Yun 2004, private communication).  {\it Spitzer} detections with 
IRAC at 3.6, 4.5, 5.8, and 8.0 $\mu$m exist for all three SCUBA sources, as well as 24 $\mu$m detections with MIPS for SCUBA sources 1 and 8.  
(Three IRAC and MIPS sources are seen within a radius of 8\arcsec\ of SCUBA source 8.)  Because of the 31\arcsec\ size of the Bolocam beam, we are 
likely influenced by source confusion.

\subsection {Bolocam Nondetections}

The Bolocam nondetections are as follows:

\noindent {\it SCUBA.LE850.14, 18.---}These galaxy candidates have SCUBA and MAMBO detections, with {\it Spitzer} IRAC and MIPS and VLA 
(Ivison et al.\ 2002, M.\ Yun 2004, private communication) counterparts.  SCUBA source 14 is discernible in the Bolocam observations at 3.9 mJy, just below 
the 
4.2 mJy, 3 $\sigma$ detection threshold.  The Bolocam pixel coincident with SCUBA source 18 has 
a flux density of 1.6 mJy, well below the detection threshold.

\noindent {\it SCUBA.LE850.7, 35.---}These sources are detected by SCUBA, {\it Spitzer} IRAC and MIPS, and VLA (7, Ivison et al.\ 2002; 7, 35 M.\ Yun 2004, 
private communication; 7, Ciliegi et al.\ 2003; 35, Egami et al.\ 2004).  The flux density in the Bolocam map coinciding with SCUBA source 7 is 3.4 
mJy, below the 3 $\sigma$ detection threshold.  At a 
Bolocam flux 
density of 0.9 mJy, SCUBA source 35 is well into the Bolocam noise.

\section{Number Counts}
\label{section:simulations}


In this section we discuss the extraction of the number (per unit
flux density per unit solid angle) versus flux density relation
(``number counts'') from the observed sources.  Because of the
presence of noise (due to the instrument, the atmosphere, and confused
background sources), there is a bias in both the observed flux
densities and the observed histogram of number of sources versus
flux density.  This bias, first noted by \cite{eddington13,eddington40}, is quite generic when attempting to
measure a statistical distribution in the presence of noise.  Further,
because our S/N with respect to these noise sources is not
large, this bias is an effect comparable in size to the statistical
Poisson errors in determining the number counts.

There are two broad approaches to extracting the number counts in the
regime where bias is significant.  The first is to directly correct
the observed number versus flux histogram using some knowledge of the
statistics of the survey.  The other approach is to assume a model and
attempt to match its parameters to the data using a fit, aided by
simulation.  The direct correction approach does not appear promising
for this survey.  \citet{eddington40} showed that, in the presence of
Gaussian measurement noise, one could apply an asymptotic series
correction to the observed distribution to obtain a better estimate of
the underlying distribution.  This correction involves even-numbered
derivatives of the observed distribution, and so, with our observed
distribution containing only 17 sources, this method is impractical.
Another approach might be to individually correct each source by its
expected bias, but \citet{eddington40} also showed that using the
distribution of corrected fluxes as a measure of the underlying
distribution is fundamentally incorrect.  Thus, we have elected to fit
a model to the data.

The formalism for relating a given underlying number count
distribution to the observed number counts is given in \S\
\ref{subsec:Definitions}.  This provides the definition of the survey
bias, completeness, and false detection rate.  The simulations used to
determine these quantities are described in \S\
\ref{subsection:noisemaps}, and their actual calculation is given in
\S\ \ref{subsec:surveycalculations}.  The effect of confusion
noise on the survey is discussed in \S\
\ref{subsec:ConfusionNoise}.  The method of extracting the underlying
counts is given in \S\ \ref{subsection:modeldnc}.  Caveats and
difficulties in extracting the underlying number counts, as well as
suggestions for improvements in a future analysis, are discussed in
\S\ \ref{subsec:SysEffectsNumCounts}.

\subsection{Formalism}
\label{subsec:Definitions}

For a given observing frequency band, we denote the differential number
count (DNC) distribution of galaxies per unit flux density interval per
solid angle as $N'(S)$.
%
The cumulative number count (CNC)
distribution will be denoted $N(S)$,
with units of number per solid angle.  The relation between the true
$N'$ and the observed distribution $n'$ must account for the effects
of random noise, the presence of a detection threshold, and confusion
noise (i.e., a contribution to the variance of the map due to sources
below the detection threshold).  As a result of all forms of noise, a source
having flux density $S$ is in general observed with a different flux
density $s$. Let $B(s, S; N')$ be the probability density that a source
with true flux density $S$ is observed at a flux density $s$; the
implicit dependence on the confusion noise is included by the
parametric dependence on $N'$.  $B(s, S; N')$ is normalized such that
\begin{eqnarray}
\label{eq:BNorm}
\int_{-\infty}^\infty{B(s, S; N') \D s} = 1
\end{eqnarray}
for all values of $S$. 
%
The quantity $B(s, S; N')$ will be referred to as the ``survey
bias''.  By normalizing according to equation (\ref{eq:BNorm}), one
assumes that a source of true flux $S$ will be found at some flux $s$
with probability unity.  In the presence of a detection threshold,
however, sources whose observed flux fluctuates below the threshold
will not be included in $n'$. In this case, the integral in equation
(\ref{eq:BNorm}) is not 1, but $C(S; N')$, the ``survey
completeness,'' namely, the probability that the source is found at all.
Note that this also depends on the confusion noise through $N'$.  
In addition, there may be {\it false detections} of noise
fluctuations, $F(s)$, which contribute to the observed number counts.  Thus, the
expression for the observed DNC distribution is
\begin{equation}
\label{eq:finalDNC}
n'(s) = F(s) + \int_0^\infty{ B(s, S; N') C(S; N') N' \D S}.
\end{equation}
In the following, the dependence of $B$ and $C$ on $N'$ will not be
written explicitly.

Under the assumptions of uniform Gaussian noise with rms $\sigma$,
negligible contribution from confusion noise, and a fixed detection
threshold $n\sigma$, analytical expressions for $C(S)$, $B(s, S)$, and
$F(s)$ can be derived.  For future reference, these are
\begin{equation}
\label{eq:GaussianCompleteness}
C(S) = \frac{1}{\sqrt{2 \pi} \sigma}
\int_{m\sigma}^\infty{ \exp{\left[\frac{-(s-S)^2}{2 \sigma^2}\right]}\D s},
\end{equation}
\begin{equation}
\label{eq:GaussianBias}
B(s, S) = \frac{1}{C(S)} \frac{1}{\sqrt{2 \pi} \sigma}
\exp{\left[\frac{-(s-S)^2}{2 \sigma^2}\right]} \Theta(s - m\sigma),
\end{equation}
\begin{equation}
\label{eq:GaussianFalseDetections}
F(s) = \frac{{\cal N}}{\sqrt{2 \pi} \sigma}
\exp{\left[\frac{-s^2}{2 \sigma^2}\right]} \Theta(s - m\sigma),
\end{equation}
where $\Theta(x)$ is the unit step function ($\Theta = 1$ for $x > 0$,
$\Theta = 0$ for $x < 0$) and ${\cal N}$ is a normalization factor for ${\cal N}$ independent noise 
elements.  

\subsection{Simulation of Noise Maps}
\label{subsection:noisemaps}

Two types of simulations were done to determine the survey bias,
completeness, and false detection rate.  Both methods simulate only
the instrument and atmospheric noise (the ``random noise'') and do not
include the effect of confusion noise.  This is appropriate for the
case in which the random noise dominates.  The validity of this
assumption is discussed in \S\ \ref{subsec:ConfusionNoise}.

In the first suite of simulations, the observational data were used to
generate 100 fake maps (noise realizations) by jittering the
individual time streams 60\arcsec\ in right ascension/declination
coordinates with a random phase before they were co-added to make maps.
This had the effect of washing out the point sources as discussed in
\S\ \ref{subsection:nulltests}.  Note that realizations of these
jittered maps are not fully independent because the noise is somewhat
correlated between realizations; the average correlation coefficient
between maps is 4\%.  Statistical error bars on the completeness and
bias determined from this simulation method include the contribution
from the correlation.  The pointing jitter dilutes the variance of
sources present in the jittered map to 20\% or less of its value in
the unjittered map (see Fig.\ \ref{fig:jitter}), effectively removing
confusion noise.

Because of the large amount of time required to generate many
realizations of maps from real time stream data (as in the case of the
jittered maps) and the difficulties of fully simulating time stream
realizations of instrument and atmospheric noise, a second simulation
method was developed.  In this method, the noise properties are
derived from the jackknife maps, which represent realizations of
signal-free instrument noise.  The noise model for the map (before
optimal filtering) assumes that the noise map, $n(\vect{x})$, can be
described as an independent noise per pixel that scales as
$1/\sqrt{t_i}$, where $t_i$ is the integration time in pixel $i$,
combined with a mild pixel-to-pixel correlation.  This correlation is
assumed to be stationary over the map and can thus be described by
the two-dimensional PSD of the noise map $\xi^2(\vect{k})$, normalized
so that its integral has unit variance.
%
%
These assumptions are justified because, as shown in \S\
\ref{section:sourcelist}, the coverage variation accounts for most of
the point-to-point variation in the noise, and examination of the
jackknife map PSD shows that the $1/k$ contribution to the PSD is
small compared to the white term, leading to largely uncorrelated
pixels.  The noise model for the map after Wiener filtering is
straightforwardly obtained by convolving the noise map with the Wiener
filter.

The assumptions above are equivalent to writing the covariance matrix
$\vect{C}$ for the unfiltered map as
\begin{displaymath}
\vect{C} = \vect{D}^{1/2} {\cal F}^{-1} \xi^2 {\cal F} \vect{D}^{1/2},
\end{displaymath}
where $\vect{D}$ is diagonal in pixel space and describes the coverage
variations, $\xi$ is diagonal in $\vect{k}$-space and describes the
pixel noise correlations, and ${\cal F}$ is the discrete Fourier
transform.  The elements of $\vect{D}$ can be written as
%
%
\begin{displaymath}
D^{1/2}_{ij} = \sqrt{ \frac{A}{t_i}} \sigma \delta_{ij}.
\end{displaymath}
where $\sigma^2$ is the sample variance of the noise and $A$ is a
normalization that ensures that the sum of the pixel variances $\sum_i
A\sigma^2/t_i$ is equal to $(N-1)\sigma^2$, the total noise variance
in the map.
%
%
The noise map should satisfy $<n n^T> = \vect{C}$; a given
realization is
\begin{displaymath}
n = \vect{D}^{1/2} {\cal F}^{-1} \xi {\cal F} w,
\end{displaymath}
where $w$ is a realization of uncorrelated, Gaussian,
mean zero, unit variance noise.
Determining the noise model then reduces to determining the form of
$\xi(\vect{k})$ and the value of $\sigma$.  The PSD $\xi^2$ is
computed directly from the uniform coverage region of the unfiltered
jackknife maps using the discrete Fourier transform; multiple
jackknife realizations (which are nearly independent) and adjacent
$\vect{k}$-space bins are averaged to reduce the noise on the
measurement of the PSD.  The overall noise normalization $\sigma$ is
determined by requiring that the variance of $n(\vect{x})$, when
considered in the good coverage region, Wiener filtered, and
multiplied by $\sqrt{t_i}$, is equal to the variance similarly
determined from the jackknife maps (\S\ \ref{subsection:nulltests}
and Fig.\ \ref{fig:jackknife}).  One thousand noise realizations were
generated in this way.

\subsection{Calculations of False Detection Rate, Bias, and Completeness}
\label{subsec:surveycalculations}

\begin{figure}
\epsscale{1.0}
\plotone{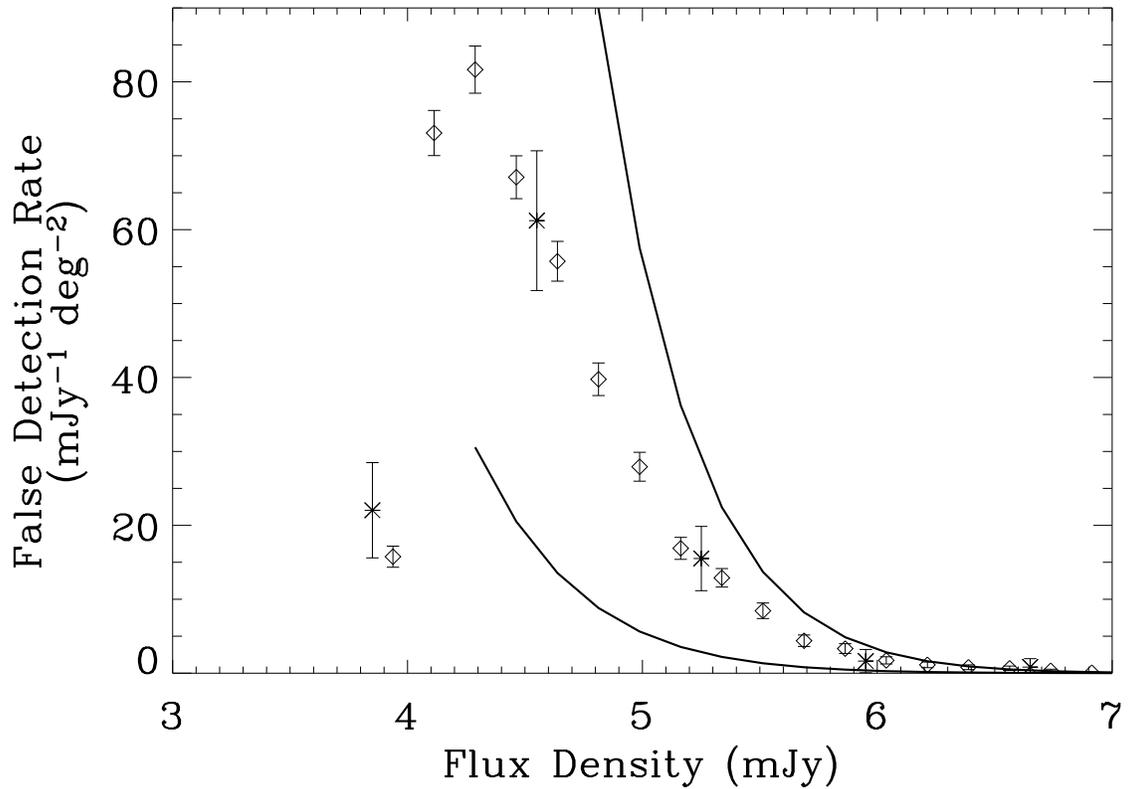}
\caption{False detection rate plotted vs.\ flux density.  Diamonds are simulations from the 
map statistics, and asterisks are simulations from the data, shown with error bars for the finite number of 
simulation realizations.  The two solid curves show the theoretical bounds on this quantity, the lower curve 
assuming that the number of independent statistical elements is the number of beams in the map, and the upper 
assuming that it is the number of pixels in the map.}
\label{fig:falsedetections}
\end{figure}

The false detections were determined by simply running the source
detection algorithm on each of the simulated maps for both types of
simulations and recording the number and recovered flux density of
the detections.  Figure \ref{fig:falsedetections} shows the results
for both methods.  Also plotted is the theoretical prediction,
assuming that the normalization ${\cal N}$ is either $N_{\mathrm{beams}}$ or
$N_{\mathrm{pixels}}$, which should bracket the possibilities.  It is seen that
neither Gaussian model describes the simulated false detection rate
well, although both methods of simulation agree well with each other.
The Gaussian model does not describe the simulation data well in
either amplitude or shape.  The amplitude discrepancy occurs
because ${\cal N}$ is the number of effective independent noise
elements, which depends on both the correlations in the noise and the
detection algorithm, which does not count all pixels above threshold
as source candidates but considers all pixels within a merged group
to be a single source.  The shape of the Gaussian model fits poorly
owing to three effects: First, because of the coverage variation, the
threshold is not sharply defined in flux density units, causing some
false detections {\em below} the threshold.  Second, the grouping
algorithm merges closely spaced false detections in the
Wiener-filtered map and assigns a single flux density to the brightest
pixel, a conditional probability that is flux dependent.  Finally, the
pixels are not independent, since both $1/f$ noise and the Wiener
filter correlate them.  Because of the difficulty in deriving an
analytic expression for all these effects, the false detection rate as
determined by simulation is used in further analysis (see \S\
\ref{subsection:modeldnc}) instead of the Gaussian prediction.  The
mean number of false detections in the Lockman Hole map as determined from simulation is 6
(Poisson distributed).

To find the completeness and bias, sources of known flux density were
injected into the noise maps.  First, a source-only map was created by
adding 30 31\arcsec\ FWHM two-dimensional Gaussians at a specified flux density
level
to a blank map.  The sources were injected at random into the uniform
coverage region but were spaced far enough apart that the source
detection algorithm could distinguish each of them; this
circumvented potential complications involving source confusion.
Then, the source-only map was added to a noise map to simulate a sky
map post cleaning and mapping.  Next, this map was Wiener filtered and
run through the source extraction algorithm, which enabled us to
determine which sources were detected and their resulting flux
densities.  Each extracted detection was centroided to determine its
position, and then its position was compared with the position of the
nearest injected source.  If the positions were within 15\arcsec\
(roughly the distance between two adjacent, diagonal pixels), the
extracted detection was considered real.  The flux density was
determined by the maximum value of detected pixels, as is appropriate
for the Wiener-filtering algorithm.

\begin{figure}
\epsscale{1.0}
\plotone{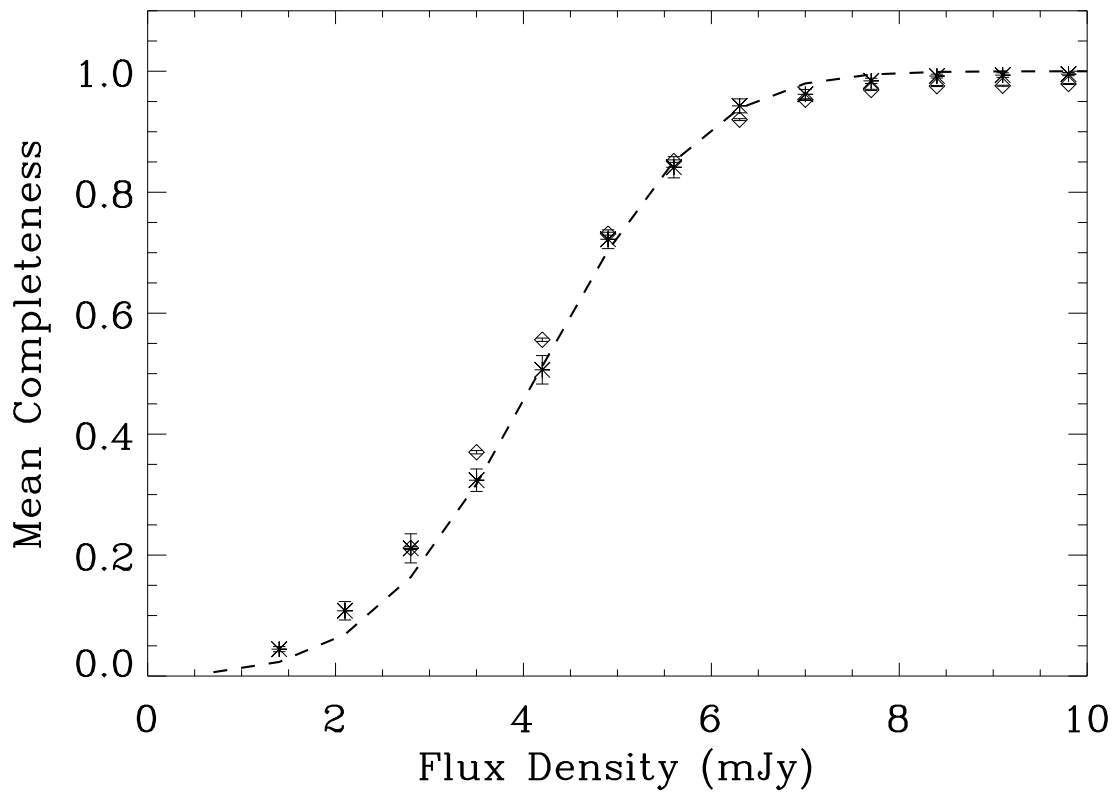}
\caption{The survey completeness as a function of flux density. Diamonds are simulations from the map
 statistics, and asterisks are simulations from the jittered data, shown with error bars for the finite number of simulation
 realizations.  The dashed curve shows the theoretical prediction for the Gaussian case with $\sigma = 1.4$~mJy (the mean noise level of the cleaned Lockman Hole map).  The error bars are statistical only and do 
 not reflect systematic differences in the two simulation methods.}
\label{fig:completeness}
\end{figure}

With these mechanics in place, the completeness was calculated by
computing the ratio of the number of detections at a given flux to the
number of injected sources.  This was repeated for source flux
densities ranging from 2.8 to 9.8 mJy for simulations from map statistics and from 1.4 to 9.8 mJy for jittered data simulations in 0.7 mJy 
intervals, with the results plotted in Figure
\ref{fig:completeness}.  The two types of simulations agree well.  The
survey completeness is 50\% at the 3 $\sigma$ detection threshold, as
expected, because half of the sources at the threshold will be bumped
upward by noise and half will be bumped downward.  The simulations
also agree with the theoretical prediction for Gaussian noise.

\begin{figure}
\epsscale{1.0}
\plotone{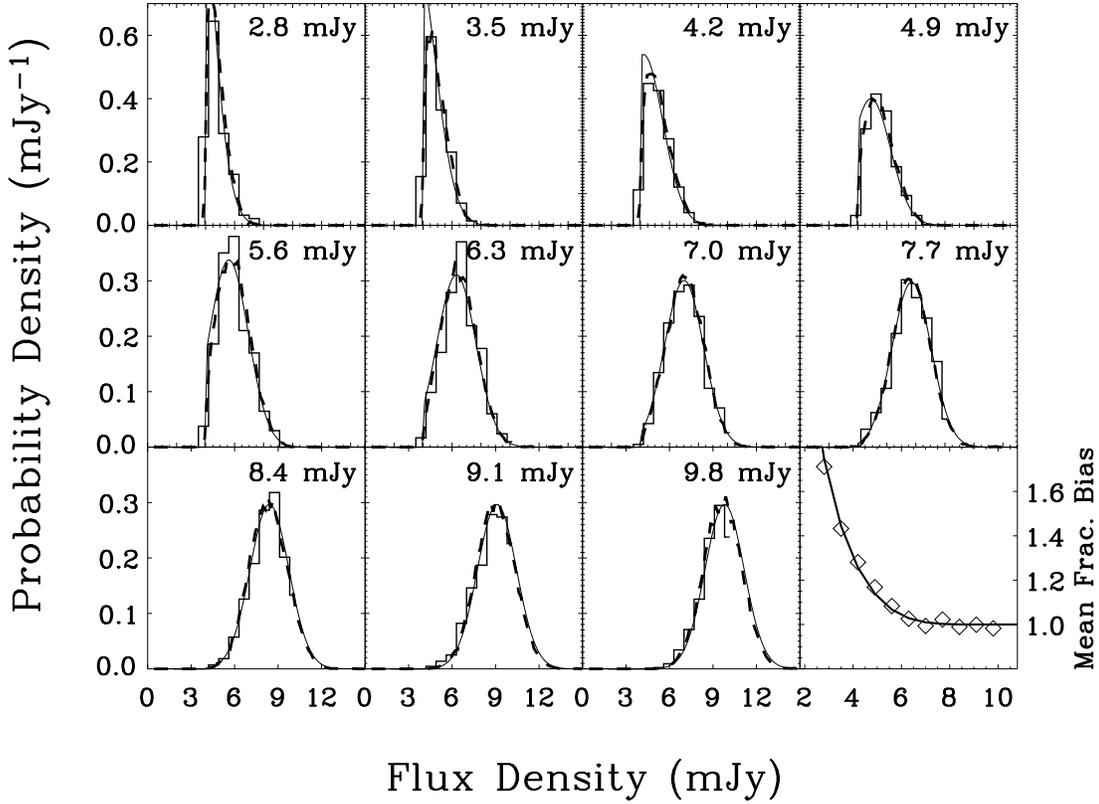}
\caption{Survey bias as a function of observed flux density for a variety of input flux densities.  The bias is normalized as a
 probability density as a function of observed flux density.  The flux densities of the sources injected in the simulation are shown in the upper 
right of each panel.  The solid histograms are simulations from the jittered data, the dashed curves are simulations 
from the map statistics, and the solid curves (which lie nearly on top of the map statistics
 simulation) are the Gaussian predictions for $\sigma = 1.4$ mJy (the mean noise level of the cleaned Lockman Hole map).  The bottom right panel shows the mean 
fractional bias as a function of input flux density for the simulations from the map statistics.  The solid curve is the 
Gaussian prediction for $\sigma = 1.4$~mJy.}
\label{fig:bias}
\end{figure}

The bias was computed by determining the distribution of measured flux
densities as a function of injected flux densities.  At relatively
large flux densities, the bias distribution should approach a Gaussian
distribution centered at the injected flux density, with $\sigma$
equal to the map rms.  This is seen to be the case in Figure
\ref{fig:bias} for injected flux densities $\ge7$ mJy.  The figure
gives the distribution of the expected observed flux densities
(probability density per flux bin, normalized to an integral of unity)
for a range of injected flux densities.  For low injected flux
densities approaching the detection threshold, the distributions
become increasingly asymmetric owing to the presence of the threshold.
The distributions do not drop abruptly to zero below the threshold
because there are variations in the map coverage.  Note that sources
with true flux below the detection threshold may be detected.  The
average bias for a source is shown in Figure \ref{fig:bias}; this
rises steeply for sources with fluxes near or below the detection
threshold.

%
The preceding discussion (in particular the agreement of the
simulated bias and completeness with the Gaussian theoretical
prediction) indicates that, in spite of coverage variations and
correlated noise, the noise in this survey behaves substantially like
uniform Gaussian noise.  Comparison of the results of the map
simulation method with the jitter technique also shows good agreement,
indicating that the assumptions that went into the map simulation
method are justified and that we have a reasonable model for the
survey noise.  This gives added confidence to the determination of the
false detection rate, which depends only on the noise properties.

\subsection{Effects of Confusion Noise on the Bias and Completeness Functions}
\label{subsec:ConfusionNoise}

The completeness and bias function estimates as determined in \S\S\
\ref{subsection:noisemaps} and \ref{subsec:surveycalculations} do not
include the effects of confusion noise.  The effect of confusion noise
is illustrated by considering two extremes: instrument noise dominant
over confusion noise and vice versa.  When instrument noise is
dominant, the bias function for this survey is correctly described by
equation (\ref{eq:GaussianBias}).  In the confusion-dominated limit, the
bias function takes on the shape of the source count distribution,
reflecting the fact that it is the underlying distribution of sources
that may bias the flux of a given source.  In between these two
extremes, the Gaussian bias function acquires additional width and a
long positive tail from the source counts distribution.  This tail
increases the probability that a low flux source will fluctuate above
the detection threshold.  Consequently, the completeness at low flux
densities is increased over the case of Gaussian noise.  Note that
small changes in the bias tail can cause large changes in the
subthreshold completeness.  The case at hand falls in this in-between
regime.  Understanding the modification to the bias function by
confusion noise is necessary for accurately estimating how confusion
transforms a model source count distribution to an observed one, as in
equation (\ref{eq:finalDNC}).  It is difficult to precisely model the
effects of confusion on bias and completeness because they depend on
the source count distribution that one is trying to measure.

We can estimate the size of the confusion noise present in our maps by
finding the relative contributions of the noise and signal variances.
The sample (per pixel) variance of the optimally filtered Lockman map
in the good coverage region is found to be 2.37~mJy$^2$.  The variance
of the optimally filtered jackknife maps in the same region is
1.81~mJy$^2$, leaving 0.56~mJy$^2$ due to sources.  
The variance contributed by all the sources in Table
\ref{table:sourcelist} is approximately 0.33~mJy$^2$, of which
0.10~mJy$^2$ is expected to be due to false detections of random noise
peaks.
This leaves 0.33~mJy$^2$ due to undetected sources.  This represents an
S/N per pixel of 0.37 in rms units; considered in
quadrature with the 1.81~mJy$^2$ of the noise, it increases the noise
estimates and the rms of the bias function by about 9\%.  

To estimate the effect of confusion noise on the survey completeness
and bias, particularly in the tail, sources were injected one at a
time into the real map and extracted using the source extraction
algorithm, with the completeness and bias calculated as in the
noise-only case.  This has the effect of making only a small change in
the observed distribution of pixel values, effectively preserving that
distribution.  No effort was taken to avoid the positions of source
candidates, as this would bias the procedure by failing to take into
account the tail of the distribution. 
%
%
This test showed that the bias acquired a high flux density tail, as
expected, and the completeness was increased above its Gaussian noise
value.  It should be emphasized, however, that this method provides an
upper limit because it effectively ``double-counts'' confusion: the
map into which the sources are injected is already confused.
Positions of high flux in the true map may already consist of two
coincident lower flux sources, and so the probability of a third source
lying on top of them is not truly as high as the probability we would
calculate by this procedure.  The determination of the completeness
and bias in this way is also limited by the statistics of only having
one realization of the confusion noise.
Applying these new bias and completeness functions, as well as the
Gaussian noise-only bias and completeness (eqs.\
[\ref{eq:GaussianCompleteness}] and [\ref{eq:GaussianBias}]), to a power-law model of the number counts (the best-fit model of \S\
\ref{subsection:modeldnc}),
the change in the observed counts is of order the size 
of the 68\% confidence interval for Poisson errors in the observed counts.
Thus, confusion noise is not wholly negligible nor does it dominate.
In extracting the number counts, we ignore the confusion noise but
discuss how to treat it correctly in \S\
\ref{subsec:SysEffectsNumCounts}.

\subsection{Fitting a Model to the Differential Number Counts}
\label{subsection:modeldnc}

To extract number counts from this data set, we use equation
(\ref{eq:finalDNC}) with the simulation-derived false detections and the
completeness and bias of equations (\ref{eq:GaussianCompleteness}) and
(\ref{eq:GaussianBias}).  A model for $N'$ is also required.  Because of the
small number of detections, the model must have as few free parameters
as possible so that the data will be able to constrain the model
parameters.  This pushes us away from detailed, physically motivated
models and toward a simple model in combination with several,
somewhat arbitrary, constraints.  We use a two-parameter power-law model
for $N'$ given by
\begin{equation}
\label{eq:dnds_model}
N'(S;\vect{p}) = A \left(\frac{S_0}{S}\right)^\delta,
\end{equation}
where $\vect{p} = [A, \delta]$ and $S_0$ is a fixed constant (not a
parameter of the model).  The choice of this form for $S_0 \neq 1$
reduces the degeneracy between $A$ and $\delta$ that prevails over
narrow ranges of $S$, such as in this survey.  We have set $S_0 =
4$~mJy.

The unaltered model of equation (\ref{eq:dnds_model}) is unsatisfactory
at both high and low flux values.  At low fluxes, the model diverges,
requiring a cutoff on which the result depends.  The issue of the low-flux cutoff is discussed further in \S\
\ref{subsec:SysEffectsNumCounts}.  For now, we simply impose a low-flux cutoff $S_l = 1$~mJy in the integral over $S$ in equation
(\ref{eq:finalDNC}).  In addition, if the model is extended indefinitely to
high fluxes, it may produce too many sources to be consistent with the
lack of observed sources.  This constraint nevertheless does not
determine the shape of number counts above the highest flux observed.
Thus, one must either implement a high-flux cutoff or assume something
about the shape of the number counts beyond the region where they are
measured.  To address this, a single bin of the same width as the
other bins has been added to the data at high fluxes, where the data
are zero and the model nonzero; beyond $S_h = 7.4$~mJy, the model is
zero.  Fixing the upper cutoff as above and allowing the lower to
float to its best-fit value produces $S_l = 1.3$~mJy.  Two additional
possibilities were also tried for a high-flux cutoff: (1) setting the
model to zero beyond the highest filled bin resulted in a very shallow
index ($\delta < 2$), and (2) allowing the highest bin to extend to
infinity produced a very steep power law ($\delta > 10$).  While both
of these cases are unphysical, they illustrate the sensitivity of
the power-law model on the high-flux cutoff.  Thus, the constraints
that have been adopted, while arbitrary, serve to restrict the range
of possible models sufficiently to extract reasonable values of
$[A, \delta]$.  However, in light of this arbitrariness, the resulting
constraint on the parameters of the power-law model must be treated
with skepticism.

To fit to the model, the data are first binned.  The number of sources
with observed flux between $s_k$ and $s_{k+1}$ is denoted by $n_k$.  We
assume that the number of sources counted in any interval $ds$ follows
an approximate Poisson process and therefore that each $n_k$ is a
Poisson-distributed random variable that is independent of $n_j$ for
$k \neq j$.  The same would not be true of the cumulative counts, and
so the differential counts are preferred for this analysis.  The 
likelihood of observing the data $\{n_k\}$ if the model is $\{N_k\}$
is then
\begin{equation}
\label{eq:likelihood}
{\cal L} = \prod_k \frac{N_k^{n_k} \exp{[-N_k]}}{n_k!}
\end{equation}
because it is assumed that the bins are independent.  The value of the
model in a given observed bin is defined as
\begin{equation}
\label{eq:model}
N_k(\vect{p}) =
\frac{1}{\Delta s} \int_{s_k}^{s_{k+1}}{\left(F(s) +
\int_0^\infty{ B(s,S) C(S) N'(S;\vect{p}) \D S}\right) \D s}.
\end{equation}
The function $-\ln{\cal L}$ is minimized with respect to $\vect{p}$ to
find the maximum likelihood value of $\vect{p}$.
%

%
Two modifications of the likelihood equation (\ref{eq:likelihood}) were
made for this analysis.  The first is that a prior was applied
to constrain $\delta > 2$, so that both the integral of the number
counts and the integral of the total flux density remained finite for $S >
0$.
Thus,
\begin{eqnarray}
\nonumber
{\cal L'} = {\cal L} \Theta(\delta - 2).
\end{eqnarray}
Second, to extract confidence regions for the fitted
parameters, it was necessary to normalize ${\cal L'}$, such that
\begin{eqnarray}
\nonumber
\int{{\cal L'(\vect{p})} \D \vect{p}} = 1.
\end{eqnarray}
This normalization was done by numerical evaluation of the likelihood
and its integral over the region where it is appreciably nonzero (see Fig.\ \ref{fig:likelihood} below).

\begin{figure}[htb]
\epsscale{1.0}
\plotone{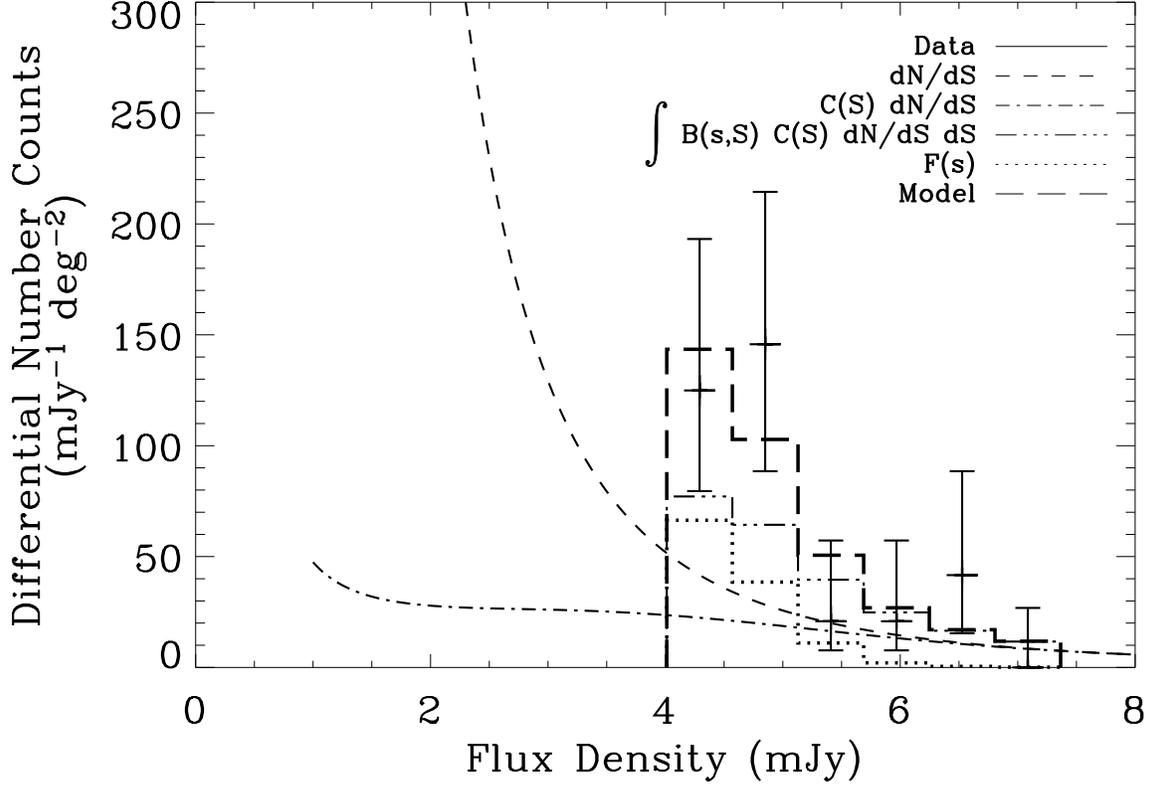}
\caption{Model of the submillimeter number count distribution as a function of flux density (results of finding the maximum 
likelihood value of the model eq.\ [\ref{eq:dnds_model}]).  The functions $N'$ and $C(S) N'$ are shown as continuous functions of true flux density, $S$,
and the data, false detection rate, $\int{B(s,S) C(S) (N') \D S}$, and model are shown binned in the
coarse bins of observed flux density, $s$, used for the raw data.  (Note that the abscissa is used for two different flux densities, $S$ and $s$.)$\:$  The ordinate was scaled from the 324 
arcmin$^2$ of the 
survey to 1 deg$^2$.  The error bars on the data are Poisson errors as described in the text.}
\label{fig:fit_results}
\end{figure}

The various components of this fit are shown in Figure
\ref{fig:fit_results}.  The data are shown with 68\% confidence
interval error bars, based on the observed number of sources in each
bin, scaled to an area of a square degree.  The error bars were
computed according to the prescription of \citet{feldman98} for small
number Poisson statistics (which unifies the treatment of upper
confidence limits and two-sided confidence intervals).  The error bar
on the highest flux density bin is an upper limit.  The model is
clearly consistent with the data given the error bars.  (All six model
bins falling within the 68\% confidence interval error bars of the
data {\em may} imply that the errors have been overestimated, although
this has a 10\% probability of occurring.) \ Examining the
fit in stages, one finds that the product of the survey completeness
and the best-fit number counts shows that the survey incompleteness
reduces the number of sources observed at low flux densities as
expected; above $\sim$7 mJy, the survey is essentially complete.  The
effect of the bias, however, combined with the steepness of the number
counts, increases the number of sources observed in all bins
substantially above that of the underlying source distribution and
contributes to the observed number of sources in all bins.  In fact,
based on the best-fit DNC and computing over the range of fluxes
observed, 67\% of real sources will have a true flux density {\em
below} the detection threshold.  Note that the best-fit number counts
lie below the Poisson errors for the raw counts, demonstrating again
that the survey bias is a nonnegligible effect.  Given the maximum
likelihood values of $A$ and $\delta$ (52.0 and 3.16, respectively),
the cumulative source count at $S_{1.1\mathrm{mm}} > 2.75$ mJy is
192$^{+108}_{-88}$ deg$^{-2}$.  This is consistent with the 1.2 mm
MAMBO number count result (378$^{+136}_{-113}$ deg$^{-2}$) for the
combined Lockman Hole and European Large-Area ISO Survey (ELAIS) N2 regions (Greve at al.\ 2004).

Contours of the likelihood function for this fit are shown in Figure
\ref{fig:likelihood}.  In calculating the likelihood, the upper and
lower flux limits were assumed to be a correct model, and as such, the
likelihood does not account for violations of this assumption.  The
shaded region was obtained by integrating the normalized likelihood
${\cal L'}$ for values ${\cal L'} > {\cal L}_{thresh}$, such that the
integral was equal to 0.68.  These are Bayesian errors that
incorporate the prior belief that $\delta > 2$.  


\subsection{Difficulties and Caveats}
\label{subsec:SysEffectsNumCounts}

As the above discussion indicates, the extraction of number counts
from this data set is subject to a number of difficulties and caveats.
In addition to the small-number statistics and the difficulty in
modeling the dependence of the survey bias and completeness on the
confusion noise that have already been discussed, a separate
discussion of the principal limitations of the preceding analysis is in
order. These are the low S/N of the detections and the sensitivity of
the result to the lowest flux objects assumed to contribute to the
observed counts.  These limitations may be overcome with a more
sophisticated future analysis, as we discuss.

The survey bias, combined with an underlying number counts
distribution rising at low fluxes, has a strong effect on the fluxes
of sources observed at low S/N.  This problem is exacerbated in the
presence of confusion noise, but it is present in surveys in which
random noise dominates over the confusion noise as well.  This point
has been appreciated in the historical confusion literature.
\citet{crawford70} showed how to use a maximum likelihood method to
extract a power-law slope from observed flux densities, and
\cite{murdoch73} extended this to the case of sources
observed with Gaussian noise.  Because of the divergence at low fluxes
of a power law, a lower limit in flux must be imposed in order to
obtain finite answers.  The principal conclusion of \cite{murdoch73} was that the
power-law slope of the number counts determined by the maximum
likelihood method depends sensitively on this lower cutoff if the S/N
of the sources used in the survey is less than 5, whereas above this
point the slope determination, while biased, is not dependent on the
lower cutoff.  This sensitivity to the lower flux cutoff also applies
to the amplitude of the power law as well, a point that is clearly
described by \citet{marshall85}.  Although the \cite{murdoch73} result is for the
rather unphysical case of a power law with an abrupt cutoff, the
general result that derived number counts based on low-S/N sources
will depend sensitively on the assumed behavior of the underlying
number counts far below threshold is more general.  This may be seen
by considering the behavior of $C(S) N'$ as $S \to 0$.  As long as
this function is increasing, the bias will continue to push some
sources of low intrinsic flux up above the detection threshold, and so
the low-S/N regime will contain sources from well below threshold.
[Note that this is consistent with the behavior of $B(s,S)$, since for
$s$ greater than threshold, $B$ is positive for all values of $S$.]
One can see from Figure \ref{fig:completeness} that the completeness
drops off rather slowly for Gaussian noise (and even more slowly when
confusion noise is added) and never vanishes.  In fact, the product
of the Gaussian completeness times any $S^{-\delta}$, $\delta > 0$,
diverges as $S \to 0$, so for many otherwise reasonable number counts
models, this problem will occur.  Thus, in the presence of any sort of
bias, whether due to confusion or noise, deriving accurate counts
above threshold requires a nontrivial amount of information about the
counts well below threshold if the S/N of the detections is low.
Since all of our sources have S/N~$\le$~5, any constraint placed on the
power-law amplitude and slope will likely depend sensitively on the
lower cutoff chosen.

An analysis technique that overcomes all of the shortcomings
mentioned above is the so-called fluctuation or ``$P(D)$'' analysis
\citep{scheuer57,scheuer74,condon74}.  This analysis matches the
shape of the observed pixel value distribution (Fig.\
\ref{fig:histogram}) against the prediction for a model distribution
combined with the instrument noise.  It overcomes the small-number
statistics by using the full map instead of only the above-threshold
detections.  It addresses the nonlinear aspects of equation
(\ref{eq:finalDNC}) by directly including the confusion noise.  Low S/N
is no longer an issue, since individual sources are no longer
considered, and the lowest flux considered is naturally at the
confusion limit where additional sources contribute only a mean to the
observed histogram.  The unchopped scan strategy of the Lockman Hole
data simplifies such an analysis because of the simplicity of the
effective beam.  Additional effects such as the attenuation of large
angular scale structure by the atmospheric cleaning or the angular
correlation of sources may also be straightforwardly included.  A
paper on this analysis is in preparation.

\section{Discussion}
\label{section:discussion}

\subsection{Comparison with Previous Number Count Results}
\label{subsection:comparenumbercounts}

A number of other groups have previously published number counts of
submillimeter galaxies.  Figure \ref{fig:number_counts} shows selected
recent results.  These include surveys of blank-fields by SCUBA on the
JCMT at 850~$\mu$m \citep{barger99, borys03, scott02}, observations
of galaxies lensed by clusters, also using SCUBA at 850~$\mu$m
\citep{blain99, chapman02, cowie02, smail02}, and blank-field surveys
by MAMBO on the IRAM telescope \citep{greve04}.  The Bolocam result is
plotted as the maximum likelihood cumulative number counts (computed
from the DNC described in \S\ \ref{subsection:modeldnc}),
evaluated from 1 mJy to the maximum observed flux density (6.8 mJy).
The number counts in the figure are not adjusted for the wavelength
differences of the surveys.

The Bolocam result is in broad agreement with previous measurements;
the maximum likelihood cumulative number counts are consistent with the
1200 $\mu$m measurement and below the 850 $\mu$m measurements, as
expected if the same population of objects is being measured.  The
region of 68\% probability in parameter space has been translated to
cumulative number counts and is shown by the region between the dashed
curves.  This region does not correspond to the naive expectation of
Poisson errors based on the number of detected sources.  This is due
to both the strong effect of the bias and the flux cutoffs imposed
on the model.  Because of the bias, it is inappropriate to assume that
the number of observed sources in a bin can be used as a measure of
the uncertainty in the underlying number counts in that bin.  The
effect of assuming an upper flux cutoff is particularly evident in the
figure in the rapid drop of the cumulative counts as the cutoff is
approached.  This causes the error bars to be artificially small, as
any model is constrained to be zero beyond this point.  The maximum likelihood number count model presented 
here, as well as its errors, depends strongly on the exact low- and high-flux cutoffs assumed for the underlying distribution and
consequently cannot be treated as definitive. 

In addition to the caveats above, it should be borne in mind that the
uncertainty of the flux bias discussed in \S\
\ref{subsection:calibration} (derived from the rms pointing error between
the Bolocam galaxy candidates and coincident radio sources) introduces a
systematic shift in the simple model of equation (\ref{eq:dnds_model}):
the parameter $A$ changes by an amount $(1 \pm \sigma_\epsilon /
\epsilon)^{-\delta}$.  This gives a steep dependence of the amplitude of
the number counts on the calibration error and the presumed power-law
index.

At high flux densities, where the survey is nearly complete and the
effect of the bias is smallest, model-independent constraints may be
obtained.  In particular, the lack of any observed sources with flux
density greater than 8 mJy has been used to place a 90\% upper
confidence limit on the cumulative number counts above 8 mJy; this is
shown by the dashed horizontal line in the figure.  This constraint
depends only linearly on the calibration error.  Bolocam appears to be
measuring near the region where the number counts, based on both the
850 and 1200 $\mu\mathrm{m}$ measurements, would be
expected to turn over, but because of the limited survey area, we do
not strongly constrain the number counts at the bright end of the
luminosity function.

\subsection{Integrated Flux Density}
\label{subsec:IntegratedFluxDensity}

The fraction of the FIRAS integrated far-infrared background light
\citep{fixsen98} measured by this survey can be computed in several
ways.  Summing the flux densities of all observed sources gives 85
mJy, or 3.9\% of the FIRAS background over the survey area at 1.1 mm
($8.0 \times 10^{-22}$ W m$^{-2}$ sr$^{-1}$ Hz$^{-1}$).  Subtracting
out the expected mean flux of false detections gives 58 mJy, or 2.7\%
of the FIRAS background.  Integrating the maximum likelihood DNC
between 1 and 6.8 mJy (the maximum observed) gives 276 mJy, or
$\sim 13$\% of the FIRAS background.  Since it seems plausible that
the number counts do in fact steepen beyond the upper range of our
observations, we conclude that at least $\sim$95\% of the light from
submillimeter sources lies below the detection threshold of this
survey and $\sim87$\% below the minimum flux derived from our number
counts model.

\subsection{Implied Luminosities and Star Formation Rates}
\label{subsection:luminosities}

The flux density of a galaxy at an observed frequency, $\nu$, is related to its intrinsic luminosity, $L$, by \citep{blain02}
\begin{eqnarray}
\label{eqn:luminosity}
S_\nu=\frac{1+z}{4 \pi D_L^2}L \frac{f_{\nu(1+z)}}{\int{f_{\nu'} \D \nu'}},
\end{eqnarray}
where $D_L$ is the luminosity distance to redshift $z$, $f_{\nu(1+z)}$ is the redshifted SED of the galaxy, and $\int{f_{\nu'} \D 
\nu'}$ is the integrated rest SED.
For a flat ($\Omega_k=0$) cosmology, it can be shown \citep[e.g.,][]{peebles93} that the luminosity distance is given by
\begin{eqnarray}
\nonumber
D_L = \frac{c\ (1+z)}{H_0}\int_0^z{\frac{1}{\Omega_M(1+z')^3+\Omega_\Lambda}\D z'}.
\end{eqnarray}

To estimate the bolometric luminosities of the submillimeter galaxies detected by Bolocam, a template SED was constructed that 
assumes a blackbody emission spectrum modified by a dust emissivity term:
\begin{eqnarray}
\label{eqn:opticaldepth}
f_\nu \propto \epsilon_\nu B_\nu(T) \propto [1-\exp{(-\tau_\nu)}] B_\nu(T),
\end{eqnarray}
where $B_\nu(T)$ is the Planck function evaluated at dust temperature, $T$, and frequency, $\nu$, and $\tau_\nu$ is the optical depth of the dust:
\begin{eqnarray}
\nonumber
\tau_\nu = \left(\frac{\nu}{\nu_0}\right)^\beta.
\end{eqnarray}  

The dust emissivity index, $\beta$, is believed to lie between 1 and 2 \citep{dunne00}.  The form of equation 
(\ref{eqn:opticaldepth}) is commonly assumed in 
the literature for dusty nearby galaxies and high-redshift AGNs, 
including \cite{benford99}, \cite{omont01}, \cite{priddey01}, and \cite{isaak02}.  This equation reduces to a simple optically thin emission 
spectrum, $\epsilon_\nu B_\nu(T) \sim \nu^{2+\beta}$, in the Rayleigh-Jeans limit and $\nu \ll \nu_0$, and it asymptotes 
to $B_\nu(T)$ at high frequencies (because an emissivity of $> 1$ is unphysical).  Observations of luminous low-redshift galaxies (Arp 220 and Mrk 231) and high-redshift galaxies detected by 
deep submillimeter surveys furthermore suggest that a power law, $f_\nu \propto \nu^\alpha$, is appropriate to model the hotter components of dust 
on the Wien side of the spectrum \citep{blain99}.  We implement such a power law at high frequencies, matched to 
equation (\ref{eqn:opticaldepth}) at 1.2$\nu_0$.


Creating a composite SED of nearby dusty {\it IRAS} galaxies, high-redshift submillimeter galaxies, gravitationally lensed high-redshift 
galaxies, and high-redshift AGNs \citep[][and references therein]{blain02}, we find that parameters of $T$ = 40 K, $\nu_0$ = 3700 GHz, $\beta$ = 1.6, and $\alpha$ = 
-1.7 provide a 
reasonable fit.  Assuming a cosmology of $\Omega_\lambda$ = 0.7, $\Omega_M$ = 0.3, and $h_0$ = 
0.73 and a galaxy redshift of $z$ = 2.4 (the median redshift that Chapman et al.\ [2003b] derive for their sample of 10 
submillimeter galaxies identified using high-resolution radio observations), equation 
(\ref{eqn:luminosity}) gives extreme bolometric luminosities of $L = (1.0-1.6)
\times 10^{13}$ $\mathrm{L}_\odot$ for the range of flux densities detected by Bolocam.  The derived luminosities are insensitive to redshift, varying by less than 25\% for $0.6 < z < 12$.  
Assuming dust temperatures of 
30 and 50 K implies luminosities of $(3.5-5.9) \times 10^{12}$ $\mathrm{L}_\odot$ and $(1.8-3.0) \times 10^{13}$ $\mathrm{L}_\odot$, respectively.  If these galaxies are lensed, their 
intrinsic luminosities will be 
lower.  

Observations of nearby star-forming galaxies suggest the following relation between the SFR present in a galaxy and 
its far-infrared luminosity:
\begin{eqnarray}
\nonumber
\mathrm{SFR}=\epsilon \times 10^{-10} \frac{L_{60 \mu \mathrm{m}}}{\mathrm{L}_\odot}\mathrm{M_\odot}\mathrm{yr}^{-1},
\end{eqnarray}
where $L_{60 \mu \mathrm{m}}$ is the 60 $\mu$m luminosity.  The value of $\epsilon$ varies in the literature from 2.1 to 6.5 \citep{scoville83, thronson86, 
rowan97} because of different assumptions about the duration of 
the starburst, different initial mass functions (IMFs), and lower mass limits.  In this paper we adopt a value of 
$\epsilon=2.1$ from a ''cirrus'' model that combines very small grains 
and polycyclic aromatic 
hydrocarbons (PAHs) with a Salpeter IMF in a starburst of OBA stars over $2 \times 10^{6}$ yr \citep{thronson86}.  Obtaining $L_{60 \mu \mathrm{m}}$ from our model SED yields large SFRs of 
$480-810$ 
$\mathrm{M_\odot}\mathrm{yr}^{-1}$.  While these calculated luminosities and SFRs are sensitive to the SED model 
parameters, particularly $T$ and $\beta$, most recent models of 
local star-forming galaxies nevertheless result in dust temperatures and emissivities that imply extreme luminosities 
and SFRs.  It is possible, however, that these extremely luminous galaxies derive some 
of their power from AGNs \citep[e.g.,][]{alexander03}, in which case the SFRs have been overestimated.  Observations of ultraluminous infrared galaxies (ULIRGs) in the local universe ($z 
\lesssim 
0.1$) with luminosities $> 10^{13}$ $\mathrm{L}_\odot$ show that nearly all of these galaxies possess luminous AGNs and 
that the dominant power source in the majority of nearby ULIRGs may be AGNs rather than star formation (Sanders 1999).  
Recent X-ray observations and optical spectroscopic data of $z > 1$ ultraluminous galaxies, however, indicate that in 
almost all 
cases the AGNs account for $< 20\%$ of the total bolometric output of higher-redshift galaxies \citep{alexander04}.  {\it Spitzer} observations will prove useful in investigating the 
incidence of AGNs versus star formation in submillimeter galaxies from the shape of the 
mid-infrared continuum emission; initial results confirm the 
high-redshift X-ray results and show a mixture of infrared-warm AGNs and cooler starburst-dominated sources \citep{egami04, frayer04}, with a smaller fraction ($\sim$25\%) of energetically 
important AGNs \citep{ivison04}.

\section {Future Work}
\label{section:futurework}

Observations at shorter submillimeter wavelengths are vital to both confirm the Bolocam galaxy candidate detections and make photometric
redshift and temperature estimates.  Follow-up 350 $\mu$m photometry of the Bolocam-detected submillimeter galaxies with SHARC-II is planned to fill in the SEDs of these galaxies.  
Precise astrometry afforded by the radio identifications (as well as the 10\arcsec\ beam size of SHARC-II) will allow optical and infrared counterparts to be identified.  Furthermore, {\it Spitzer} 
far-infrared observations combined with the Bolocam 1.1 mm galaxy survey will provide a flux
density ratio that is strongly dependent on redshift for a given temperature.  This is because the rest wavelength 
corresponding to the observed {\it Spitzer} wavelength
of 70 $\mu$m is on the rapidly falling Wien side of the greybody spectrum (for a $z \sim 2$ galaxy at 40 K), 
and Bolocam's 1.1 mm observations are on the steep
$\nu^{2+\beta}$ ($\beta\approx$1.5) modified Rayleigh-Jeans side of the SED.  The ratio of 
$S_{1.1\mathrm{mm}}/S_{70\mu\mathrm{m}}$ is thus highly dependent on redshift, growing by a 
factor of 250 from $z=1$ to 5.  {\it Spitzer} IRAC observations will also provide independent photometric redshift 
determinations from the SEDs of stellar populations of submillimeter galaxies redshifted into the near-IR.  
These combined observations, in conjunction with the radio-to-far-infrared correlation \citep{yun02}, will thus 
allow the temperature and redshift distributions of these  
submillimeter galaxies to be constrained.  A detailed discussion of the SEDs and photometric redshift/temperature 
estimates 
of the Lockman Hole galaxies will follow in a companion paper (in preparation).

As discussed in \S\ \ref{subsection:comparenumbercounts}, this survey does not constrain the number counts at flux densities above 7 mJy, at approximately
the break point where the number counts are expected to drop sharply (based on the 850 $\mu$m observations in Fig.\ \ref{fig:number_counts}).  
This can be addressed with a survey covering a larger area to shallower
depth.  Such a survey has been started with Bolocam in the COSMOS
field,\footnote{See http://www.astro.caltech.edu/cosmos.} which
currently covers $\sim 1000$ arcmin$^2$.  This survey should allow either determination or a strong upper limit on the 1.1
mm number counts beyond 7 mJy, as well as uncover extremely bright, interesting sources, 
perhaps with strong AGN components that should be easy to follow up at other wavelengths.  In
addition to the wide area, the COSMOS observations have extremely
uniform coverage ($<$3\% rms) and a highly cross linked scan strategy that
aids in rejecting atmospheric $1/f$ noise better than the Lockman Hole observations.  


\section{Conclusions}
\label{section:conclusions}

Bolocam is a new bolometer camera with a large field of view and a rapid
mapping speed optimized for surveys, including surveys for rare, bright
submillimeter galaxies.  We have used Bolocam on the Caltech Submillimeter
Observatory at a wavelength of 1.1 mm to conduct a survey of 324 arcmin$^2$ toward the Lockman Hole for submillimeter galaxies.  Unlike previous 
submillimeter surveys, the observations were made 
without chopping, at a fast scan rate of 60\arcsec\ s$^{-1}$.  The Bolocam survey encompasses the entire 850 
$\mu$m 8 mJy JCMT SCUBA and 1.2 mm IRAM MAMBO surveys to a comparable depth under the assumption of a model SED for a galaxy at $z=2.4$, with relative rms of 1\,:\,0.9\,:\,0.6$-$1.4, 
respectively.  We 
have reduced the resulting data set using a 
custom IDL-based software pipeline, in which correlated atmospheric and instrument noise is rigorously removed via 
a PCA sky subtraction technique.  We detect 17 galaxies at a significance of $\ge3\ \sigma$, where the 
map rms is $\sim$1.4 mJy.  

A series of simulations have allowed us to verify the robustness of the galaxy candidates.  Extensive jackknife and pointing jitter tests reveal that the sources detected in this survey 
have a small characteristic length scale (point sources) and are contributed to by the ensemble of observations, strongly indicating that 
the galaxy candidates are real.  Simulations of the observations using both synthetic maps and observational data indicate that six false detections
should be expected.  

Comparing our detections to those of other surveys (including SCUBA 850 $\mu$m, MAMBO 1.2 mm, and VLA
radio observations) indicates that the majority of Bolocam sources have coincident detections in at least two other wavebands; 
we conclude that a majority of the Bolocam detections are real.  Six of the detections are galaxies previously detected by the SCUBA 8 mJy survey.  Of 
the remaining 11 Bolocam detections, 9 of them lie 
outside the SCUBA survey region, and we cannot search for counterparts for them.  Seven of the 17 Bolocam detections have been detected 
by the MAMBO 1.2 mm survey, with 6 of the remaining 10 sources lying outside the MAMBO good coverage region.  While both the SCUBA and MAMBO surveys detect most of the Bolocam sources 
in the overlap region, neither Bolocam nor SCUBA/MAMBO detect the majority of the remaining SCUBA and MAMBO sources.  A total of 65\% of the 17 
Bolocam source candidates have at least one radio coincidence, although the accidental radio detection rate is high (34\%) owing to the size 
(31\arcsec\ FWHM) of the Bolocam beam.  Furthermore, we statistically detect the aggregate average of the SCUBA and MAMBO sources below our 3 $\sigma$ detection threshold at significances 
of 3.3 and 4.0 $\sigma$, respectively.  

Further simulations enabled us to estimate the completeness and bias of this survey, which were subsequently used with the false detection rate to fit a simple power-law model of the underlying 
parent distribution to match the observed number count distribution.  This model constrains the submillimeter counts over the flux density range $S_{1.1\mathrm{mm}}$ = 1$-$7 mJy.  While the validity of this model is 
significantly limited by both the effects of confusion noise and the flux density cutoffs assumed for the underlying number count distribution, we find this 
modeled number count distribution to be consistent with previously published submillimeter galaxy number counts.  Integrating the maximum likelihood 
differential number counts distribution between 1 and 6.8 mJy (the maximum observed flux density) yields 
276 mJy in the map, or $\sim$13\% of the FIRAS integrated far-infrared background light.  

If the Bolocam galaxy candidates lie at redshifts $z>1$, then their inferred luminosities are $L = (1.0 - 
1.6) \times 10^{13}$ $\mathrm{L}_\odot$ (assuming a dust temperature of 40 K).  Further assuming that they are powered by
star formation, large SFRs of $480-810$ $\mathrm{M_\odot}\mathrm{yr}^{-1}$ are implied.  Multiwavelength 
follow-up observations of the Lockman Hole field are underway with {\it Spitzer} and SHARC-II in order to constrain the 
temperature/redshift distributions of these sources.

\acknowledgments

With alacrity, we acknowledge the support of the CSO director and staff, the support of Kathy Deniston, and
helpful conversations with Andrew Blain, Steven Eales, and Min Yun.  This
work was supported in part by NSF grants AST-0098737, AST-9980846, and AST-0206158 and
PPARC grants PPA/Y/S/2000/00101 and PPA/G/O/2002/00015.  D. H.
acknowledges the support of a PPARC Ph.D. Fellowship, S. R. G. acknowledges Caltech for the R. A. Millikan Fellowship, and G. T. L.
acknowledges NASA for GSRP Fellowship NGT5-50384.

\clearpage

\clearpage

\clearpage

\clearpage

\clearpage

\clearpage

\clearpage

\clearpage

\clearpage

\clearpage

\clearpage

\clearpage

\clearpage

\clearpage
\begin{figure}[htb]
\epsscale{1.0}
\plotone{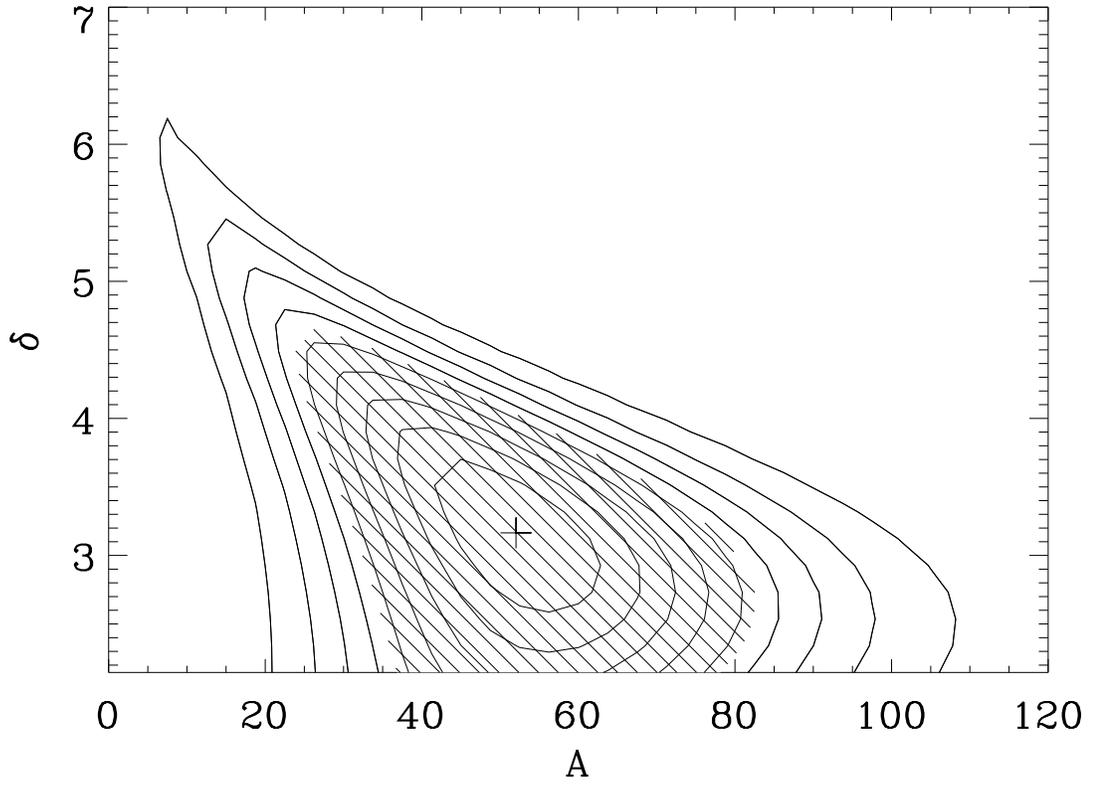}
\caption{Contours of the likelihood function (eq.\
[\ref{eq:likelihood}]) in the $[A, \delta]$-plane.  Contours are shown
for 10\%, 20\%, \ldots, 90\% of the peak height.  The location of the peak
is shown by a cross, and the hatched area indicates the 68\%
confidence region.  The maximum likelihood values are $A = 52.0$ and
$\delta = 3.16$.}
\label{fig:likelihood}
\end{figure}

\clearpage
\begin{figure}[htb]
\epsscale{1.0}
\plotone{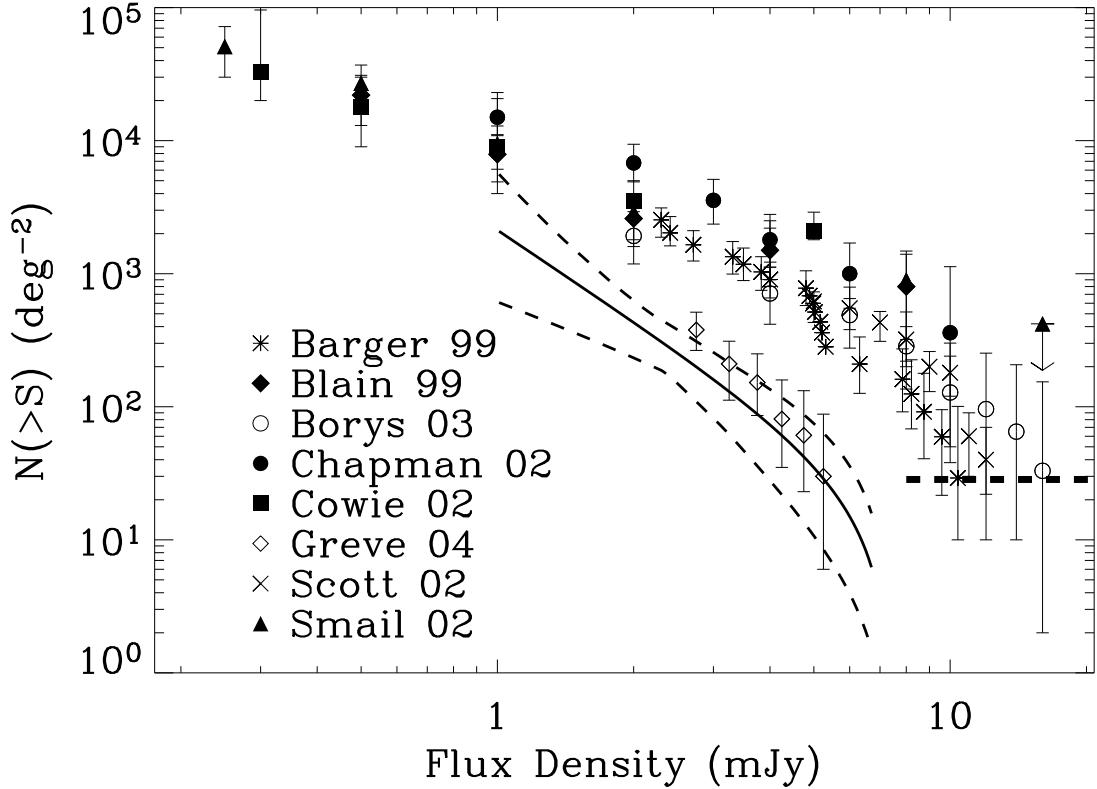}
\caption{Previous measurements of the number counts of submillimeter
galaxies, along with the new measurement with Bolocam.  Previous
surveys that used lensing by clusters are indicated by filled
symbols; all other symbols indicate blank-field surveys.  The solid
line shows the integral of the Bolocam maximum likelihood differential number
counts; the region between the dashed curves is the 68\% confidence region of Fig.\
\ref{fig:likelihood} translated into the range of possible cumulative
number counts.  The thick horizontal dashed line is the 90\% upper confidence limit for
sources brighter than 8 mJy for this survey.  All surveys are at 850 $\mu\mathrm{m}$, except for
Bolocam at 1100 $\mu\mathrm{m}$ and \cite{greve04} (IRAM MAMBO) at 1200 $\mu\mathrm{m}$.  See \S\S\ \ref{subsection:modeldnc} and
\ref{subsec:SysEffectsNumCounts} for a description of the limitations of this model.}
\label{fig:number_counts}
\end{figure}

\clearpage

\clearpage

\begin{deluxetable}{lc}
\tablecaption{Observational Parameters}
\tablewidth{0pt}
\tablehead{
\colhead{Quantity} & \colhead{Value}
}
\startdata
Field of View & 8\arcmin \\
Beam Size (FWHM) & 31\arcsec \\
1.1 mm Band Center & 265 GHz \\
Bandwidth & 42 GHz \\
Raster Scan Speed\tablenotemark{a} & 60\arcsec\ s$^{-1}$ \\
Chopped Observation Scan Speed\tablenotemark{b} & 5\arcsec\ s$^{-1}$ \\
Chopper Throw\tablenotemark{c} & 90\arcsec \\
Subscan Step Size & 162\arcsec \\
Subscan Substep Size & $\pm11$\arcsec \\
\enddata
\tablenotetext{a}{Raster scans were scanned in right ascension with steps between
subscans in declination.}
\tablenotetext{b}{Chopped observations were scanned in azimuth with steps
between subscans in elevation.}
\tablenotetext{c}{For chopped observations only.}
\label{table:observationalparameters}
\end{deluxetable}

\clearpage

\begin{deluxetable}{cccccc}
\tablecaption{Galaxy Candidates}
\tablewidth{0pt}
\tablehead{
\colhead{Source} & \colhead{R.A. (J2000.0)} & \colhead{decl. (J2000.0)} &
\colhead{S/N} & \colhead{Corrected S$_\nu$} & \colhead{Corrected $\sigma$}
}
\startdata
1&10:52:55.5&57:21:03&5.0&6.8&1.4\\
2&10:51:16.7&57:16:33&4.8&6.5&1.4\\
3&10:52:12.2&57:15:53&4.5&6.0&1.4\\
4&10:52:03.6&57:18:23&4.0&5.2&1.4\\
5&10:52:29.5&57:22:03&3.9&5.1&1.3\\
6&10:51:15.6&57:14:23&3.5&5.0&1.5\\
7&10:51:30.0&57:31:03&3.4&4.9&1.5\\
8&10:52:37.0&57:24:33&3.7&4.8&1.3\\
9&10:53:05.2&57:15:23&3.1&4.8&1.5\\
10&10:51:31.4&57:23:13&3.5&4.7&1.4\\
11&10:52:48.0&57:12:43&3.2&4.6&1.5\\
12&10:51:15.5&57:15:23&3.2&4.6&1.4\\
13&10:52:35.7&57:17:53&3.3&4.5&1.4\\
14&10:52:01.1&57:25:03&3.3&4.4&1.3\\
15&10:51:47.4&57:28:33&3.2&4.4&1.4\\
16&10:52:27.1&57:25:13&3.1&4.1&1.4\\
17&10:51:59.9&57:24:23&3.1&4.0&1.3\\
\enddata
\tablecomments{Units of right ascension are hours, minutes, and seconds, and units of declination are degrees, arcminutes, and arcseconds.}
\label{table:sourcelist}
\end{deluxetable}

\clearpage

\begin{deluxetable}{ccccccc}
\tablecaption{Summary of Coincident Detections}
\tablewidth{0pt}
\tablehead{
\multicolumn{6}{c}{\hspace{3.6cm} Fraction of Galaxy Candidates Detected}\\
\colhead{Survey}&& \colhead{Bolocam} & \colhead{SCUBA} & \colhead{MAMBO} & \colhead{Radio} & \colhead{Accidental Radio}
}
\startdata
Bolocam/CSO&&-&6/8&7/11&12/17&6\\
SCUBA/JCMT&&7/31&-&8/31&15/31&3\\
MAMBO/IRAM&&8/23&8/17&-&11/23&1\\
\enddata
\tablecomments{Each row corresponds to the fraction of sources detected {\em by} each survey (i.e. 6 of 8 Bolocam galaxy candidates in the overlap region were detected by SCUBA, whereas 
Bolocam detected 7 of 31 SCUBA sources).  Column of radio detections include detections by \cite{ivison02}, \cite{ciliegi03} and/or M.\ Yun (private communication).  The 
accidental detection rate is that expected from 
a random distribution of these known radio sources within the 2 $\sigma$ confidence regions of Fig.\ \ref{fig:ancillary}.}
\label{table:ancillary}
\end{deluxetable}

\clearpage

\clearpage

\end{document}